\documentclass[prb,aps,twocolumn,tighten,floatfix]{revtex4-1}
\usepackage[pdftex]{graphicx}
\usepackage{bm}
\usepackage{color}
\usepackage{amssymb}
\usepackage{mathrsfs}
\usepackage{floatflt}

\renewcommand{\deg}{$^{\circ}$\hspace{1mm}}

\newcommand{\etal}{ {\it et al.}}
\newcommand{\newc}{\newcommand}
 
\newc{\be}{\begin{equation}}
\newc{\ee}{\end{equation}}
\newc{\bfe}{\begin{floatequation}}
\newc{\efe}{\end{floatequation}}
\newc{\bea}{\begin{eqnarray}}
\newc{\eea}{\end{eqnarray}}
\newc{\ie}{{\it i.e.} }
\newc{\eg}{{\it e.g.} }
\newc{\etc}{{\it etc.} }
\newc{\ra}{\rightarrow}
\newc{\lra}{\leftrightarrow}
\newc{\lsim}{\buildrel\langle\over{\sim}}
\newc{\gsim}{\buildrel\rangle\over{\sim}}

\newc{\one}{\mathbbm{1}}
\newc{\Tr}[1]{\mathrm{Tr}\left[ {#1} \right]}
\newc{\ket}[1]{\left|{#1}\right\rangle}
\newc{\bra}[1]{\left\langle{#1}\right|}
\newc{\braket}[2]{\langle{#1}|{#2}\rangle}
\newc{\mean}[1]{\langle{#1}\rangle}
\newc{\braketd}[1]{\langle{#1}|{#1}\rangle}
\newc{\ketbrad}[1]{\left|{#1}\rangle\!\langle{#1}\right|}
\newc{\ketbra}[2]{\left|{#1}\rangle\!\langle{#2}\right|}
\newc{\EV}[2]{\langle{#1}\rangle_{#2}}
\newc{\C}{\ensuremath{\mathbbm C}}

\def\kbt{k_{B}T}
\def\be{\begin{equation}}
\def\ee{\end{equation}}

\begin{document}
\title{Classical noise, Quantum noise and Secure communication}
\author{C. Tannous and J. Langlois} 
\affiliation{Laboratoire de Magn\'etisme de Bretagne - CNRS FRE 3117\\
UBO, C.S.93837 - 29238 Brest Cedex 3 - FRANCE}

\date{\textcolor{blue}{\today}}

\begin{abstract}
Secure communication based on message encryption might be performed by
combining the message with controlled noise (called pseudo-noise)
as performed in Spread-Spectrum communication used presently in 
Wi-Fi and Smartphone Telecommunication systems. 
Quantum communication based on entanglement is another route for 
securing communications as demonstrated
by several important experiments described in this work.
The central role played by the photon in unifying
the description of Classical and Quantum noise as major 
ingredients of secure communication systems is highlighted
and described on the basis of the classical and quantum fluctuation
dissipation theorems. 
\end{abstract}
\maketitle

\section{Introduction}

Secure communication based on message encryption with controlled noise (pseudo-noise or PN)
started with the work of the actress-engineer Hedy Lamarr and her husband-pianist 
Georges Antheil in 1941 who were interested in military communications during World War II.

Lamarr invented frequency hopping to prevent an intruder from jamming a signal sent to 
control torpedoes remotely, since using a single frequency might be easily detected and blocked.
Frequency hopping is used presently in Bluetooth and other types of wireless communication 
and is called FHSS~\cite{Carlson} (Frequency Hopping Spread Spectrum). 

Given a set of frequency values $[f_1, f_2, f_3...]$, one selects a well-defined 
sequence of frequencies following an apparently random pattern (picked from PN values) 
shared solely between transmit and receive ends. 
For an eavesdropper unaware of the sequence used, the signal appears as white noise containing no
valuable information and that is the reason why it is termed spread-spectrum communication
given that noise has broader bandwidth than the signal.

FHSS needs another important ingredient to be completely operational: sender-receiver 
perfect synchronization in order to be able to modulate-demodulate with the right frequency. 
It is Antheil, exploiting his musician skills, who developed the 
synchronization~\cite{Carlson} method between sender and 
receiver enabling them to encrypt-decrypt ongoing transmitted information.

The analog to digital conversion of frequency hopping gave birth to DSSS
(Direct Sequence Spread Spectrum) methods based on Galois polynomials, the generators of
PN sequences that we describe below and that are used presently in Wi-Fi and other 
types of digital communications (Smartphones...).


Consequently, it is important to relate noise to communications, how it might 
be used to alter the nature of the signal and ultimately transmit hidden information
in a way such that it is  properly retrieved by the target receiver.

An ordinary resistor has a fluctuating
voltage across it whether standing free or belonging to an electronic 
circuit. The resistor embodies free electrons that are thermally agitated, 
inducing random voltage fluctuations. 

Most physicists/engineers refer to this thermal voltage fluctuation as 
Johnson-Nyquist noise, after J.B. Johnson~\cite{Nyquist}, who was first 
to observe the effect at Bell laboratories in 1928 and H. Nyquist~\cite{Nyquist} 
who first explained it. 

Circuit noise studies in Bell laboratories have a very peculiar history since
in the 1950's, H. E. D. Scovil and his associates  built the world's lowest-noise
microwave amplifiers cooled by liquid helium to reduce noise  
and incorporated in extremely sensitive radiometers used in radio-astronomy.

Radiometers usually contain calibration noise sources consisting of a resistor 
at a known temperature. During the 1960's Penzias and Wilson~\cite{Penzias} 
while improving these radiometers discovered  serendipitously Big-Bang cosmic 
background radiation in 1965.


B. Yurke~\cite{Yurke} and his collaborators embarked, in the 1980's, 
on a pioneering study of Quantum noise through the quantization of $LC$ networks 
drawing from an analogy between an $LC$ circuit and the harmonic oscillator. 
Quantum effects in circuits occur when we deal with low temperature 
(as in superconductors) or at very high frequency. Usual telecommunication and 
signal processing frequencies are in the kHz-GHz range whereas 
Tera-Hz ($10^{12}$ Hz) devices encountered in medical imaging and 
optical devices operate at $10^{14}$ Hz. Consequently, kHz-GHz frequencies
are classical whereas Tera-Hz and optical devices should be considered as quantum.
With the progress of integrated circuits toward the nanometer scale (presently 
the minimal feature used in the semiconductor industry is 14 nm) and 
single electron as well as quantum dot (akin to synthetic atom) devices, 
we expect large quantum effects implying quantum noise becoming more 
important than thermal. 


The equivalence between an impedance
and an oscillator is a very important idea that will trigger and sustain steady
progress in several areas of Quantum information and communication.

Nyquist derived an expression for White Noise based on the interaction between 
electrons and electromagnetic waves propagating along a transmission line using
arguments based on black-body radiation. This means that Nyquist is in fact
a true pioneer in Quantum noise. 

He based his work on Johnson measurements who 
found that thermal agitation of electricity in conductors produces a random 
voltage variation between the ends of the conductor $R$ of the form:

\be
\langle(V - \langle V \rangle)^2 \rangle= \mean{\delta V^2}= 4R k_B T \Delta f
\ee

$\langle ...\rangle$ is the average value, voltage
fluctuation is $\delta V = V-\mean{V}$ and $V$ is the instantaneous voltage measured at
the ends of the resistance $R$. 
$k_B$ is Boltzmann constant and $T$ is absolute temperature. $ \Delta f$ is the
bandwidth of voltage fluctuations (see Appendix A).
This frequency interval spans the range of a few Hz to several tens of GHz.

The voltage fluctuation developed across the
ends of the conductor due to Thermal noise is unaffected by the presence or 
absence of direct current. This can be explained by the fact that electron thermal 
velocities in a conductor are much greater ($\sim 10^3$ times) than electron drift velocities. 

Since electromagnetic waves are equivalent to photons through Quantum Mechanics 
Duality principle~\cite{Duality}, Nyquist derivation is based on blackbody radiation that was
explained earlier by Planck.

In Quantum Mechanics language, a (zero rest mass) photon is a special case of 
a harmonic oscillator since the energy levels are separated by the same energy $\hbar \omega$ 
i.e. the $n$-th level $E_n=n\hbar \omega$ (ignoring zero-point energy $\hbar \omega/2$) 
corresponds to an integer number $n$ of photons. 
Moreover $\omega=2\pi f$ is the electromagnetic pulsation
and not the mechanical one $\sqrt{k/m}$ where $k,m$ are the respective 
spring constant and  mass of the mechanical oscillator.


While classical pseudo-noise used in spread-spectrum communications hides the signal
from intrusion by an eavesdropper through a crypting operation (FHSS frequencies follow
a PN sequence whereas in DSSS, the signal is directly multiplied by the PN sequence) using
a set of keys (corresponding to a given PN sequence) that are shared solely between 
the transmitter and the receiver, Quantum mechanics can be used to encrypt the 
signal in a completely different fashion.


Quantum Mechanics can be used to generate naturally random instead of 
deterministic pseudo-random numbers. In the early days of computing
cosmic rays or radioactive sources~\cite{Schmidt} were used for generating 
non-deterministic random numbers.
Quantum phenomena being essentially non-deterministic, 
would be able to produce truly 
random numbers and the corresponding devices are called Quantum Random Number
Generators~\cite{Schmidt} (QRNG).  

Obviously, this is not the only advantage of Quantum Mechanics since
at the Garching Max Planck Institute for Quantum Optics (MPQ) in Germany 
and the Technical University of Vienna, 
communication experiments showed that Quantum Mechanics provides entanglement 
as an alternative concept to secure information transfer between two remote sites. 

Entanglement, first introduced by Einstein (who called it "spooky 
action at a distance"), Podolsky, and Rosen~\cite{EPR}, and 
Schr\"odinger~\cite{Schrodinger} in 1935, can arise when two 
quantum systems are produced from 
a common source, e.g. when a spinless particle decays into two particles
carrying opposite spins. Such states violate a set of 
inequalities~\cite{Bell} established by J.S. Bell in 1964, implying 
that quantum theory embodies non-locality (see section IV for the mathematical
implication). Bell inequalities are the statistical measure of entanglement
and their violation can be demonstrated by measuring correlations between
quantum states.

Entangled quantum systems behave as if they can 
affect each other instantaneously, even when they are extremely far from 
each other, due to the essential non-local~\cite{Buscemi} character of 
entanglement. 

The strongest advantage of noise-based communication is that by hiding a
signal in noise, it is extremely difficult or even impossible to detect it
if the eavesdropper does not know the keys or the algorithm used between 
the transmitter and the receiver. In Quantum communication (QC), entanglement
ties together in a very stringent fashion both parties and any intrusion
attempted by an eavesdropper, when detected, triggers immediately disruption 
of communication.

This work can be taught as an application chapter 
in a general Statistical Physics course at the Graduate  
or in a specialized Graduate course 
related to applications of Quantum Mechanics and Statistical Physics
since physicists generally interested in the applications of
Quantum Mechanics and Statistical Physics are keen to expand their 
knowledge to areas of Quantum Information Processing and Communications (QIPC).

This paper is organized as follows: after reviewing several derivations
of White noise by Nyquist and others in section II, we discuss in section III
the Fluctuation Dissipation theorem and its quantum version in order 
to derive in a rigorous way, Nyquist result with modern quantum noise approach
and lay the foundations of secure communication from the classical and quantum
points of view. In section IV we apply the analysis to secure communications
with classical noise (spread-spectrum) and entanglement based Quantum information 
processing and transfer. Discussions and Conclusions are in section V.

\section{Derivations of Thermal Noise}

Nyquist work is based on phenomenological thermodynamic considerations and electric
circuit theory, including the classical equipartition theorem.
The latter is based on the physical system number of degrees of freedom.
This number~\cite{freedom} is well defined when the different contributions to system energy 
(translational, rotational, vibrational, electromagnetic...) 
are quadratic and decoupled with presence of weak interactions.
In the general case (non-quadratic energy or strong interactions), 
one evaluates the partition function in order to derive thermodynamical properties.

\subsection{Nyquist derivation of Thermal Noise}
Nyquist based his derivation on Einstein remark that many physical systems would 
exhibit Brownian motion and that Thermal noise in circuits is nothing more 
than Brownian motion of electrons due to ambient temperature.
Despite the fact one might find several strange assumptions and even flaws in 
Nyquist derivation, it remains a pioneering interesting approach since it paves 
the way to quantization of electrical circuits and noise in circuits.
Nyquist considered thermal noise in a resistor $R$ as stemming from 
electrons interacting with electromagnetic waves represented by 
a one-dimensional black-body thermal radiator. Electromagnetic waves travel 
through an ideal (lossless) one-dimensional
transmission line of length $\ell$ joining two resistances $R$.
Hence the transmission line characteristic impedance being equal to $R$ amounts 
to considering that its impedance is matched at both ends and 
that any voltage wave propagating along the line is completely absorbed by the 
end resistor $R$ without any reflection, exactly like a black-body. 

Each resistor $R$ has a thermally-fluctuating voltage at
temperature $T$ which will be transmitted down the wires 
with a current and voltage wave appearing across the other resistor.  

\begin{figure}[htbp]
  \centering
    \includegraphics[angle=0,width=50mm,clip=]{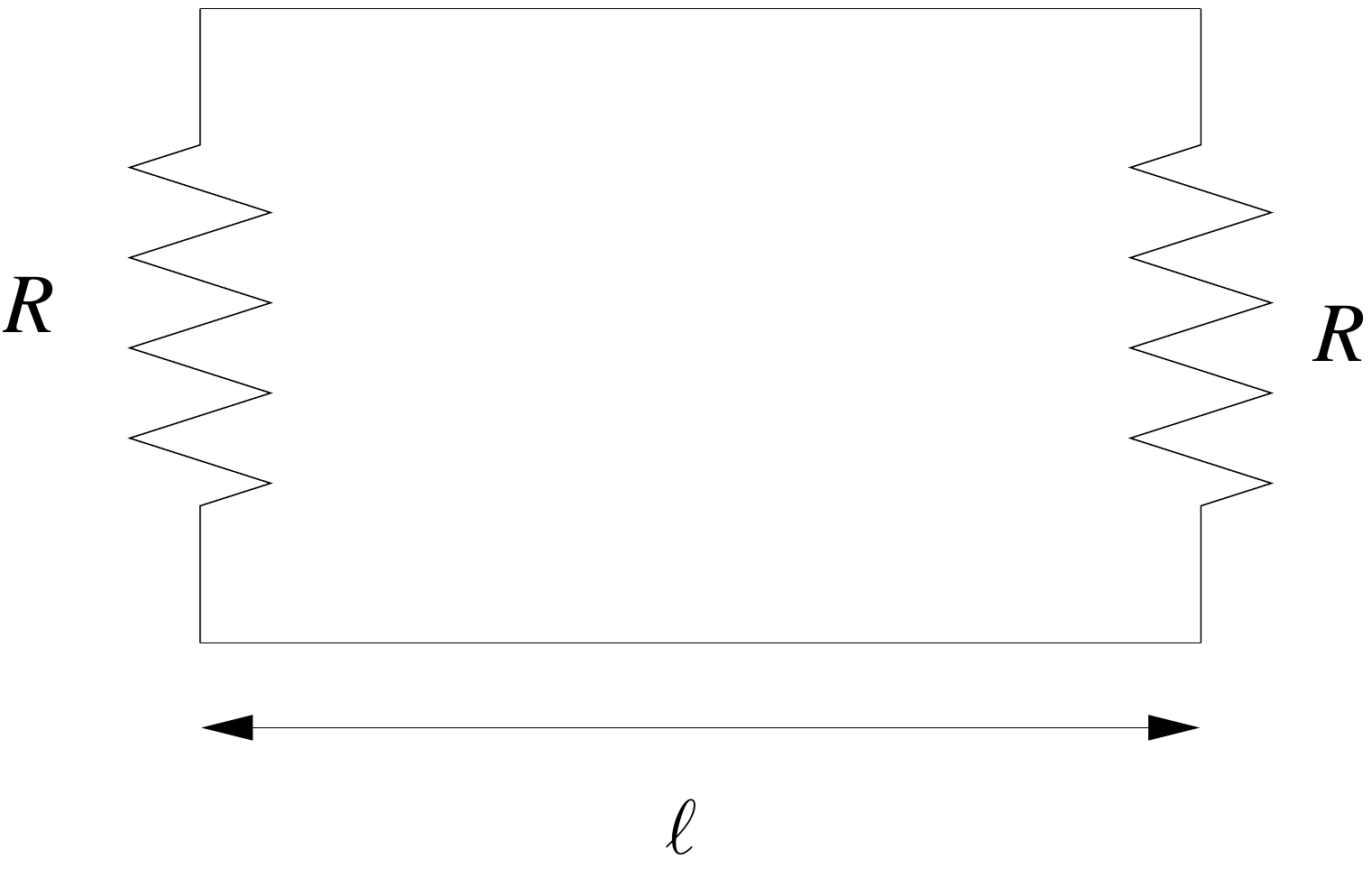}  
\vspace*{-3mm} 
\caption{Transmission line  of length $\ell$ matched by two resistors $R$ at its ends.}
    \label{fig1}
\end{figure} 

A voltage wave propagating along the transmission line is expressed at any point $x$
and any time $t$ as:
$V(x,t) = V_0 \exp[i(\kappa x - \omega t)]$  
where $\kappa$ is the wavenumber and $\omega$ the angular frequency of the wave.

The  velocity $v = \omega /\kappa$ in the line is typically $c/10$ where 
$c$ is light velocity in vacuum.
Considering the transmission line of length $\ell$ as a domain between $x=0$ and $x = \ell$ and 
imposing the boundary condition $V (\ell,t) = V (0,t), \hspace{0.5cm} \forall t$,
we infer that possible propagating wavenumbers are 
given by $\kappa \ell = 2n\pi $, where $n$ is any integer, 
and there are $\Delta n = (1/2\pi)d\kappa=d\omega/(2\pi v)$  such modes per unit 
length of the line in the frequency range between $\omega $ and $\omega  + d\omega $.

In the Canonical ensemble, the Bose-Einstein distribution for photons gives the mean number
of photons  $\langle n \rangle$  per mode at energy $\hbar \omega$ and temperature $T$ as:
\be
\langle n \rangle= \frac{1}{e^{\beta \hbar \omega} -1}
\label{Bose}
\ee

$\beta = \frac{1}{k_B T}$ is inverse temperature~\cite{Reif} coefficient. 
Thus the energy of the photon gas (ignoring zero-point energy) 
is $\langle n \rangle \hbar \omega$.

Detailed balance allows to equate the power absorbed by a resistor
(in any angular frequency range $\omega$  and $\omega  + d\omega$) 
to the power emitted by it. The energy in the interval $\omega$  and $\omega  + d\omega$
is proportional (see Appendix C) to the number of  propagating modes per 
unit length in this frequency range. The mean energy per unit time incident
upon a resistor in this frequency range is:

\be
P_{in}=v \left(\frac{d\omega}{2\pi v}\right) E(\omega)= \frac{1}{2  \pi} E(\omega) d\omega 
\ee   

where $E(\omega)$ is the electromagnetic energy at $\omega$.
Nyquist considers that the total resistance making the circulating
current $I$ is $2R$ and thus $I = V /2R$ as if the line  
whose characteristic impedance $R$ did not contribute at all to
the total resistance. Perfect matching at both ends implies that no resistance is 
contributed by the line since current/voltage waves are not subjected to scattering. 
The mean power emitted down the line and absorbed by the resistor at the other end is 
\be
R\langle I^2\rangle= R \left<\frac{V^2}{4R^2}\right>= \frac{1}{4R} 
\int\limits_0^{\infty} S_V(\omega) d\omega
\ee

where $S_V$ is the voltage Power Spectral Density (PSD) (see Appendix A).

Hence we have:

\be
\frac{1}{2  \pi} E(\omega) d\omega= \frac{1}{4R} S_V(\omega) d\omega
\ee

with:

\be
S_V(\omega) = \frac{2R}{\pi} \frac{\hbar \omega}{e^{\beta \hbar \omega} -1}
\ee

Moving from angular to linear frequency, we get: 

\be
S_V(f) = 4R\frac{hf}{e^{\beta hf} -1}
\label{frequency_dependent}
\ee

Voltage fluctuations are given by (see Appendix A):

\be
\sigma_V^2=\mean{\delta V^2}=\int\limits_{0}^{\infty}4R\frac{hf}{e^{\beta hf} -1} df
\ee

Performing a change of variable $x=\beta hf$, we get:

\be
\sigma_V^2=\frac{4R (k_B T)^2}{h} \int\limits_{0}^{\infty}\frac{x}{e^{x} -1} dx
\label{integral}
\ee

Using the integral~\cite{Landau,Gradstein}:

\be
\int\limits_{0}^{\infty}\frac{x^{2n-1}}{e^{x} -1} dx= \frac{(2\pi)^{2n}B_n}{4n}
\ee

with $B_n$ the Bernoulli polynomial coefficients, we select $n=1$ and $B_1=\frac{1}{6}$.

The value of the integral in eq.~\ref{integral} is thus $\frac{\pi^2}{6}$. 

The fluctuations are then given by:

\be
\sigma_V^2=\frac{2R (\pi k_B T)^2}{3h}
\ee

This surprising result implies that, in the classical case, 
($ h \rightarrow 0$), fluctuations become extremely large. In fact, 
ordinary frequencies $f \sim $kHz-GHz are low with respect to 
6.25 $\times 10^{12}$ Hz frequencies that correspond to 
thermal room temperature energy $k_B T$. 
Thus $hf \ll k_B T$ and the number of photons per mode is large since 
$\langle n \rangle \approx k_B T/hf$.  In the classical 
limit, the number of photons being very large, we get wave-like behaviour
whereas in the quantum limit a small $\langle n \rangle$ produces particle-like 
(photon) behaviour. 


Expanding the PSD eq.\ref{frequency_dependent} at low frequency $hf \ll k_B T$:

\be
S_V(f) \approx 4R k_B T (1 -\frac{hf}{2 k_B T})
\label{low_frequency}
\ee

Thus quantum effects no longer intervene in the low frequency limit 
$ f \rightarrow 0$, yielding Nyquist result: 

\be
S_V(0) = 4R k_B T
\label{frequency_independent}
\ee

Another divergence is encountered when we ignore the frequency dependence of 
$S_V(f)$ in eq.~\ref{frequency_dependent} and consider
$S_V(0)$ to be valid for all frequencies as usually considered for "White Noise"
(flat spectrum for all frequencies):

\be
\sigma_V^2= \int\limits_{0}^{\infty} 4R k_B T df= 4R k_B T \int\limits_{0}^{\infty} df= \infty
\ee


This divergence is similar to the Ultra-Violet catastrophe encountered in black-body
radiation since $hf \ll k_B T$ corresponds to Rayleigh-Jeans regime and its 
solution is that voltages are filtered and we never encounter 
in practice an infinite frequency domain. 

Therefore let us assume we have a finite bandwidth $\Delta f$ for voltage 
fluctuations, then:

\be
\sigma_V^2= \int\limits_{0}^{\Delta f} 4R k_B T df= 4R k_B T \Delta f
\ee

To sum up, in order to recover the Johnson-Nyquist result we have to respect two
conditions: finite band $\Delta f < \infty$ and low frequencies (kHz-GHz range) 
$\Delta f \ll k_B T/h$.

Additionally, it is surprising to note that Nyquist considered a 1D photon gas 
with a single polarization
despite the fact the photon had two polarizations (circular left and right) and in sharp 
contrast with the evaluation of the blackbody radiation by Planck who considered 
a 3D gas with two polarizations (see Appendix C). Moreover, Nyquist ignored zero-point 
energy in spite of its importance in quantum circuits 
and the fact Planck introduced it in his second paper on black-body radiation 
(see further below).

\subsection{RC circuit classical derivation of Thermal Noise}

Nyquist's theorem can be proven with the help of a parallel $RC$ circuit containing
a random source representing interactions with a thermal reservoir.
The resistor $R$ is parallel to the capacitor $C$ and the result of 
random thermal agitation of the electrons in the resistor will charge
and discharge  the capacitor in a random fashion. 

Starting from the time dependent equation of motion of the $RC$ circuit, we have:

\begin{equation}
R\frac{dq(t)}{dt}= -\frac{q(t)}{C} + \xi(t)
\label{langevin}
\end{equation}

\begin{figure}[htbp]
  \centering
    \includegraphics[angle=0,width=40mm,clip=]{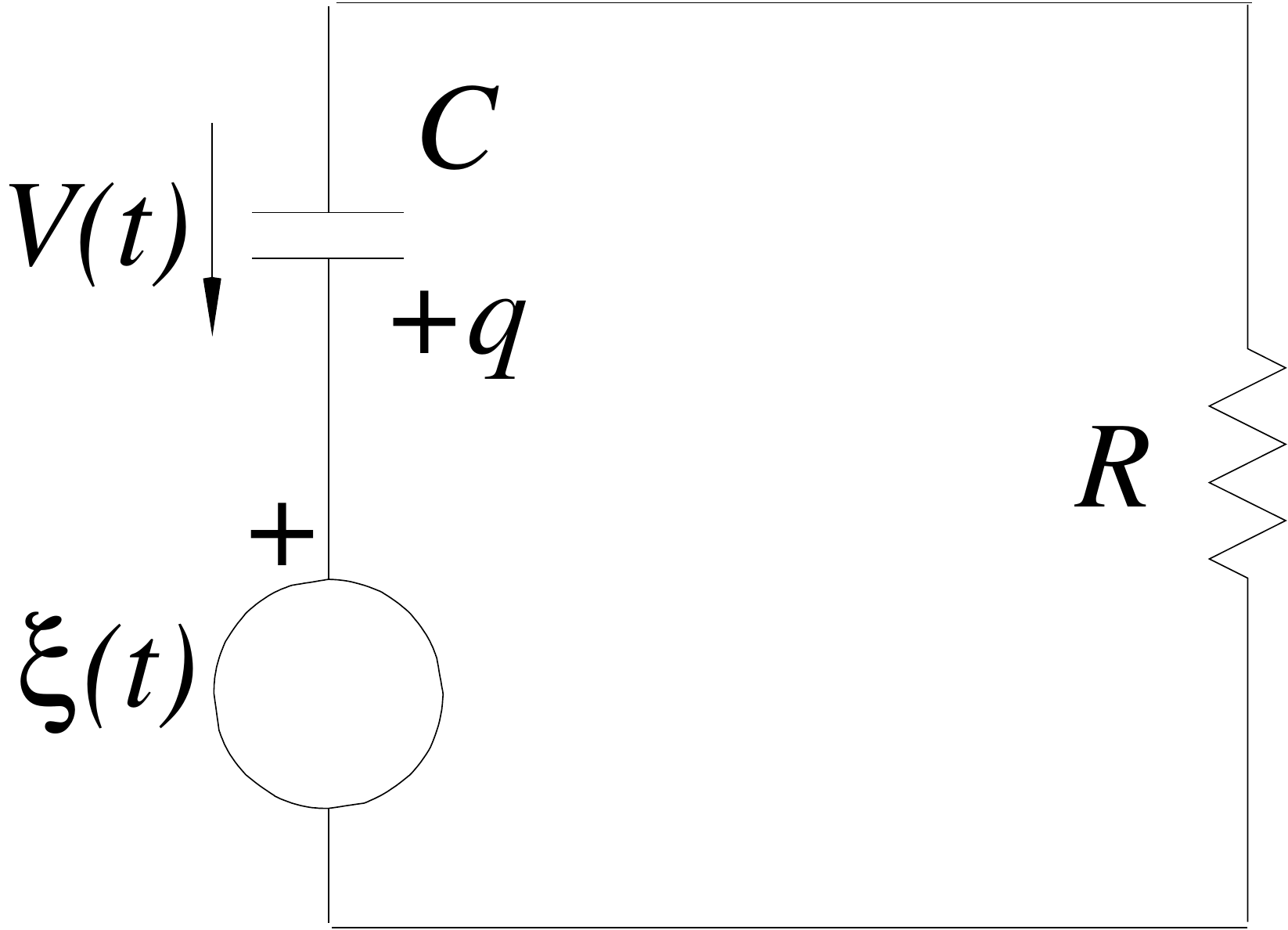}  
\caption{RC circuit with the noise source $\xi(t)$ originating from thermal contact with a reservoir.}
\vspace*{-3mm} 
    \label{RC}
\end{figure} 

$q(t)$ is the capacitor charge and $\xi(t)$ is a stochastic voltage 
(see Appendix A)  stemming from interactions with
a reservoir at temperature $T$ with the following statistical properties: 

\begin{equation}
\mean{\xi(t)}=0 \mbox{  and  } \mean{\xi(t)\xi(t')} = \lambda \delta(t-t') 
\label{white2}
\end{equation}

$\lambda$ is a constant that will be determined later and 
eq.~\ref{langevin} is called a Langevin  equation~\cite{Gardiner} (see Appendix B) due to the
presence of the time-dependent random term $\xi(t)$.

Assuming the capacitor is uncharged at time $t=-\infty$, direct integration of the 
first-order differential equation yields:

\begin{equation}
q(t)= \frac{1}{R} \exp(-t/RC) \int\limits_{-\infty}^{t}   \exp(t'/RC) \xi(t') dt'
\label{langevin2}
\end{equation}

The voltage $V(t)$ across the capacitor is related to charge through: $q(t)=C V(t)$, 
therefore evaluating the charge PSD (see Appendix A) is equivalent to voltage PSD.

Using properties of $\xi(t)$ given in eq.~\ref{white2}, we obtain:

\begin{equation}
\mean{q(t)q(t')}= C^2 \mean{V(t) V(t')}= \frac{\lambda C}{R} \exp \left(-\frac{|t-t'|}{RC}\right)
\label{spectrum}
\end{equation}

This result is expected as discussed in Appendix A since the auto-correlation
$\mean{q(t)q(t')}$ must be a decreasing function of the argument $|t-t'|$
controlled by the relaxation time $RC$. 

Setting $t=t'$ we have the equality: $C^2 \mean{V^2(t)}= \frac{\lambda C}{R}$, hence
$\lambda$ is determined as: $\lambda= RC  \mean{V^2(t)}$.

The average $\mean{V^2(t)}$ can be determined from the energy  
$\frac{1}{2}C \mean{V^2}$  stored in the capacitor
through the classical equipartition theorem: 
$\frac{1}{2}C \mean{V^2}=\frac{1}{2} k_B T$ as if a capacitor is equivalent
to a single degree of freedom (see next section).

The equipartition theorem can be proven as follows.
If a system is at temperature $T$, the probability that it is in a state of energy $E$ 
is proportional to the Boltzmann factor $\exp(-E/k_B T )$.

In the $RC$ circuit the probability element $dp$ of finding
a voltage between $V$ and $(V + dV)$ is $dp= A \exp(-E/k_B T ) dV$ corresponding 
to an energy $E=\frac{1}{2}C V^2$ stored in the capacitor $C$.

The prefactor $A$ normalizes the probability density:
\be
\int\limits_{-\infty}^{\infty} A  e^{(-C V^2/2 k_B T )} dV =1
\ee
Using the result $\int\limits_{-\infty}^{\infty} e^{-x^2} dx= \sqrt{\pi}$ we get 
$A=\sqrt{\frac{C}{2 \pi k_B T}}$.
The mean square value of the voltage is obtained from the probability density 
as:
\be
\mean{V^2}= A \int\limits_{-\infty}^{\infty}  V^2 e^{(-C V^2/2 k_B T )} dV
\ee

Thus $\mean{V^2}= \frac{k_B T}{C}$.

The voltage fluctuation PSD may be evaluated with the equipartition theorem as:

\begin{equation}
S_V (\omega)= \mean{V^2} \frac{2RC}{1+(RC \omega)^2}= \frac{2R k_B T}{1+(RC \omega)^2}
\label{spectrum2}
\end{equation}

At frequencies such that $\omega \ll 1/RC$ we get $S_V (\omega)= 2R k_B T$
recovering the Johnson-Nyquist result (see note~\cite{factor2}).

\subsection{Derivation of Thermal noise from Einstein thermodynamic fluctuation theory}
A macroscopic system at thermodynamic equilibrium is an ensemble of subsystems  which are in 
thermodynamic equilibrium with each other. The actual values of the variables, 
however, may differ from mean equilibrium values. 

The departure from equilibrium is due to fluctuations in the subsystems.

The probability, $p$, for entropy fluctuation $\delta S$ is obtained by reverting 
Boltzmann principle $S=k_B \ln W$ as $p(\delta S)=p(S-\langle S\rangle) \propto e^{\delta S/k_B}$.
The probability $p$ is proportional to $W$ the number of microscopic available states 
and we assume the validity of applying Boltzmann principle to the entropy fluctuation $\delta S$.

Generally, entropy is a function of state variables $X_i$, 
i.e. $S = S(X_1, X_2, . . ., X_i, . . .)$. It 
can be expanded as $dS$ around equilibrium since it is an analytical 
function for most thermodynamic systems when
small fluctuations of the $X_i$ are considered.

At equilibrium $S$ is maximum and all the first order derivatives 
$\frac{\partial S}{\partial X_i}=0, \forall i$
implying that the first non-vanishing terms  are quadratic.

Thus one has around equilibrium:

\be
\delta S \approx \frac{1}{2} \sum_{i,j} \frac{\partial^2 S}{\partial X_i \partial X_j} \delta X_i \delta X_j
\ee

The series expansion is performed at the equilibrium value of $X_i$, hence  
$\delta X_i =  X_i -\langle X_i\rangle$. 

The combination of Boltzmann principle and second-order expansion of entropy about equilibrium
results in a Gaussian probability density function (PDF) for finding subsystems with the non-equilibrium value of the variable $X_i$ (akin to the central limit-theorem),

\be
p(\delta X_i)= \frac{1}{\sqrt{2\pi \sigma_i^2}} e^{-\frac{\delta X_i^2}{2\sigma_i^2}}
\ee

meaning that the average value (or macroscopic equilibrium) of $X_i$ is $\langle X_i\rangle$
and that the standard deviation away from equilibrium is 
$\sigma_i^2=\mean{\delta X_i^2}=-k_B/(\frac{\partial^2 S}{\partial X_i^2})$.

The entropy of a single phase, one component system is given in terms of the energy $U$, volume 
$\Phi$ and pressure $P$ as $dS=\frac{dU}{T}+\frac{P}{T}d\Phi$.

In the presence of a voltage $V$, the entropy expression becomes 
$dS=\frac{dU}{T}+\frac{P}{T}d\Phi -\frac{q}{T}dV$ since charge $q$ couples to
voltage $V$. This yields the value 
of the entropy derivative $\left(\frac{\partial S}{\partial V}\right)_{T,\Phi}=-\frac{q}{T}$.
The second derivative is thus obtained as:
$\left(\frac{\partial^2 S}{\partial V^2}\right)_{T,\Phi}= -\frac{1}{T} \left(\frac{\partial q}{\partial V}\right)_{T,\Phi}$.

The voltage fluctuation is expressed as:

\be
\sigma_V^2=\mean{\delta V^2}=-k_B/\left(\frac{\partial^2 S}{\partial V^2}\right)_{T,\Phi}=k_B T \left(\frac{\partial V}{\partial q}\right)_{T,\Phi}
\ee 

Assuming the validity of Ohm's law: $I=\frac{dq}{dt}=\frac{V-V_0}{R}$ where 
$V_0$ is some reference voltage (e.g. $V_0= \mean{V}$), 
we infer that voltage fluctuations are given by:

\be
\sigma_V^2=k_B T R \frac{\left[d(V-V_0)/dt\right]}{(V-V_0)}
\ee

Estimating the time derivative: $\frac{d(V-V_0)}{dt} \sim \frac{(V-V_0)}{\tau}$ with
$\tau$ as a typical time variation of the voltage and given 
that for a band-limited signal of bandwidth~\cite{Carlson} $\Delta f$ with 
$\Delta f \sim \frac{1}{2 \tau}$, we finally obtain for
the voltage fluctuation expression :

\be
\sigma_V^2= 2R k_B T\Delta f
\ee

that agrees with Johnson-Nyquist result (see note~\cite{factor2}).

\section{The quantum fluctuation-dissipation theorem (QFDT)}
\label{QFDT}
The classical fluctuation-dissipation theorem derived in Appendix B provides a
relation between equilibrium fluctuations and dissipative transport
coefficients. Besides, it is an interesting route to quantize classical noise.

Callen and Welton~\cite{Callen} proved the QFDT with the correspondence theorem 
allowing to transpose classical results to quantum ones such that
a classical physical quantity is transformed into its quantum counterpart
with an observable operator.

For a single degree of freedom, linear response theory~\cite{Hanggi} yields for the
change of the expectation value of an operator-valued observable $B$ due to
the action of a (classical) force $F(t)$ that couples to the conjugate
dynamical operator $A$:

\begin{equation}
\langle\delta B(t)\rangle = \int\limits_{-\infty}^{t}ds \chi_{BA}(t-s) F(s)\,.
\end{equation}

$\delta B(t)=B(t)-\langle B\rangle_0$ denotes the difference with respect to
the thermal equilibrium average $\langle B\rangle_0$ in force absence.
The dissipative part of the response function  $\chi_{BA}(t)$ is given by:

\begin{equation}
\chi_{BA}^D(t)= \frac{1}{2i}[\chi_{BA}(t) - \chi_{AB}(-t)]\,.
\end{equation}

The fluctuations are described by the equilibrium correlation function
\begin{equation}
C_{BA}(t) = \langle\delta B(t)\delta A(0)\rangle_\beta
\end{equation}

The thermal average is taken at an inverse temperature $\beta$ (see Appendix A).

The correlation function is
complex-valued because the operators $B(t)$ and $A(0)$ in general do not
commute. While the antisymmetric part of $C_{BA}(t)$ is directly related 
to the response function by linear response theory, the symmetrized correlation function PSD:

\begin{equation}
S_{BA}(t)=\frac{1}{2}\langle\delta B(t)\delta A(0)+\delta A(0)\delta B(t)\rangle
\end{equation}

depends on the Fourier transform of the dissipative part of the response function:

\begin{equation}
S_{BA}(\omega) = \hbar\coth\left(\frac{\beta \hbar\omega}{2}\right) X_{BA}^D(\omega)\,.
\end{equation}

where $X_{BA}^D(\omega)$ is the Fourier transform of $\chi_{BA}^D(t)$. 
This is the quantum version of the fluctuation-dissipation theorem
as it links the fluctuations $S_{BA}(\omega)$ to dissipation as in the
classical case (Appendix B).

Note that $S_{BA}$ is a two variable extension of the PSD
previously used and defined in Appendix A with a single variable.
Consequently $S_V$ should be written in fact as  $S_{VV}$.

Analyzing the response of a current $\delta I$ through an 
electric circuit subject to a voltage change $\delta V$, 
implies $B=I$ and $A=Q$, since voltage couples to charge $Q$.

A circuit response is determined by $\delta I(\omega)=Y(\omega) \delta V(\omega)$ 
where $Y(\omega)$ is the admittance. 
Given $I=\dot Q$, the symmetrized current PSD is $S_{II}(\omega)=i\omega S_{IQ}(\omega)$ yielding:

\bea
S_{II} (\omega) &= \hbar\omega\coth\left(\frac{\beta \hbar\omega}{2}\right)
\Re Y(\omega)  \nonumber \\
&=2 (\mean{n}+\frac{1}{2})\hbar\omega \Re Y(\omega)\,.
\eea

where $\mean{n}$ is the Bose-Einstein factor (given by eq.~\ref{Bose}) and $\Re$ is real part symbol.
In the high temperature limit $k_B T \gg \hbar\omega$, we recover the Johnson-Nyquist result 
$ S_{II}(\omega)= 2\Re Y(\omega) k_B T$ (see note~\cite{factor2}). 

Nyquist, in the last paragraph of his 1928 paper~\cite{Nyquist},
had already anticipated the quantum case. However, he made use of
Planck first paper on black-body radiation which does not contain 
zero-point energy term $\frac{1}{2} \hbar \omega$. By missing this term, Nyquist
ignored the 1912 second~\cite{Planck} paper  on black-body radiation by Planck
who aimed at correcting his previous work by introducing zero-point energy
in order to recover the right classical~\cite{classical} limit of an oscillator 
mean energy per mode. Let us add that zero-point energy can also be shown
to originate from Heisenberg uncertainty as in the 1D harmonic oscillator 
case~\cite{Landauq}.

If the oscillator mean energy per mode is taken as $\mean{n} \hbar \omega$, 
we obtain to order $O(\frac{1}{k_B T})$ the classical (high temperature) limit $k_B T \gg \hbar\omega$:

\bea
\mean{n} \hbar \omega & \approx & \frac{\hbar \omega}{1+ \frac{\hbar \omega}{k_B T} + \frac{1}{2} {(\frac{\hbar \omega}{k_B T})}^2... -1} \nonumber \\
 & \approx & \frac{k_B T}{1+ \frac{1}{2} (\frac{\hbar \omega}{k_B T})...} \approx k_B T - \frac{1}{2} \hbar \omega.
\eea

Thus one should rather write the mean energy per mode as  $(\mean{n}+ \frac{1}{2}) \hbar \omega$ 
in order to retrieve the right classical limit $ k_B T$ originating
from the comparison between Planck photon spectral density expression 
$\frac{2 \omega^2}{\pi^2 c^3}  \left[\frac{\hbar \omega}{e^{\beta \hbar \omega} -1} \right]$
and Rayleigh-Jeans~\cite{electromag} expression $ \frac{2 \omega^2}{\pi^2 c^3} [k_B T]$
($c$ is the velocity of light).

\subsection{$LC$ circuit quantum derivation of Thermal Noise}
We start with an analogy~\cite{Devoret} between the harmonic oscillator and the $LC$ resonator. 
Moving from an $RC$ to an $LC$ circuit
stems from the fact a resistance may be defined from $ R = \sqrt{\frac{L}{C}} $ 
and that an oscillator underlying a resistor allows a ready route to quantization.

Later on when we consider a semi-infinite transmission line with $L$ inductance 
and $C$ capacitance per unit length, the line resistance
$R = \sqrt{\frac{L}{C}} $ is same as the single $LC$ resonator.
Thus the transmission line  might be viewed  simply as a large collection 
of harmonic oscillators (normal modes) and hence can be readily quantized. 
The resistance picture that links the resonator to the transmission line 
is very appealing and has has been introduced for the first time by 
Caldeira and Leggett~\cite{Devoret} to describe a continuum as sets of harmonic oscillators 
as described below.

Let us write the Hamiltonian of a single $LC$ resonator circuit in the form:

\be
\mathscr{H}_0= \frac{q^2}{2C_0} + \frac{\phi^2}{2L_0}
\ee

where variables $q$ and $\phi$ are capacitor charge and flux in the inductor.
Drawing from complete analogy with the harmonic oscillator, we quantize variables with:

\be
q=\sqrt{\frac{\hbar}{2R}}(a+a^{\dagger}), \phi=-i\sqrt{\frac{\hbar R}{2}}(a-a^{\dagger}) 
\ee

where $ R = \sqrt{\frac{L_0}{C_0}} $ and $a^{\dagger}$, $a$  are ladder operators  
characterized by commutation property: $[a,a^{\dagger}]=1$.

The Hamiltonian is transformed into standard harmonic oscillator form
$\mathscr{H}_0= \hbar \omega_0 (a^{\dagger}a + \frac{1}{2})$
with $\omega_0= 1/{\sqrt{L_0 C_0}}$, the classical resonance frequency.
 
Moving from a single oscillator to a continuum, 
we follow Goldstein~\cite{Goldstein} treatment by considering a transmission
line  of length $\ell$ characterized by an inductance $L$ and capacitance  $C$ per unit length.  

The Hamiltonian of the system is 
\be \mathscr{H}(t) = \int\limits_0^\ell dx\,\left[ \frac{q^2(x,t)}{2C} + \frac{\phi^2(x,t)}{2L}\right], 
\ee

where $\phi(x,t)$ is the local flux density and $q(x,t)$ is the local charge density.

We  define a new variable 
\be
\theta(x,t) = \int\limits_0^x dx'\, q(x',t) 
\ee 

to express current density $ j(x,t) = -\frac{\partial \theta(x,t)}{\partial t}$
and charge density $ q(x,t) = \frac{\partial \theta(x,t)}{\partial x}$
such that charge conservation rule:

\be
\frac{\partial}{\partial x} j(x,t) + \frac{\partial}{\partial t} q(x,t) = 0.
\ee 

is obeyed.

The Hamiltonian is written as:
\be \mathscr{H}(t) = \int\limits_0^\ell dx\,
\left[ \frac{1}{2C}\left(\frac{\partial \theta}{\partial x}\right)^2 
+\frac{L}{2}\left(\frac{\partial \theta}{\partial t}\right)^2 \right] 
\ee

From Hamilton equations of motion, we get the wave equation 
$\frac{\partial^2 \theta}{\partial x^2} - \frac{1}{v^2} \frac{\partial^2 \theta}{\partial t^2}= 0 $ 
with velocity $v=1/{\sqrt{LC}}$.

The normal mode expansion when the transmission line is considered with stationary
boundary conditions at both ends $\theta(0,t) = \theta(\ell,t)=0$ is given by:

\be 
\theta(x,t) = \sqrt{\frac{2}{\ell}}\sum_{n=1}^\infty b_n(t) \sin k_n x, 
\ee 

where $b_n(t)$ is the time-dependent mode amplitude and quantized wavevectors $k_n = \frac{n\pi}{\ell}$.
After substitution of this form into the Hamiltonian and integrating over $x$ exploiting orthogonality
of the basis functions $[\cos k_n x, \sin k_n x]$ over the interval $\ell$ gives: 

\be 
\mathscr{H}(t) =\sum_{n=1}^\infty  \frac{k_n^2}{2C} {[b_n(t)]}^2 + \frac{L}{2} \left[\frac{d b_n(t)}{dt}\right]^2
\ee 

Quantizing the system in terms of harmonic oscillator ladder operator sets
using the correspondence: 
\be
b_n(t) \rightarrow \sqrt{\frac{\hbar C}{2}} \frac{\sqrt{\omega_n}}{k_n}[a^\dagger_n(t) + a_n(t)]
\ee 

where $\omega_n= \frac{n v \pi}{\ell}$ controls Heisenberg time dependence of ladder operators through
$a^\dagger_n(t)=\exp(i \omega_n t) a^\dagger_n(0)$  and $a_n(t)~=~\exp(-i\omega_n t)~a_n(0)$,
yields, from charge density $ q(x,t) = \frac{\partial \theta(x,t)}{\partial x}$, the voltage at $x=0$:

\bea 
V(t) & =&\frac{1}{C}\left[\frac{\partial \theta(x,t)}{\partial x}\right]_{x=0} \nonumber \\
    & =& \sqrt{\frac{\hbar}{\ell C}}\sum_{n=1}^\infty 
\sqrt{ \omega_n} [e^{i \omega_n t} a^\dagger_n(0) + e^{-i\omega_n t}~a_n(0)] \nonumber \\
\eea

The voltage PSD is obtained after quantum averaging the voltage time correlation (see Appendix A): 
\be
S_{V}(\omega) = \frac{2 \pi}{\ell C} \sum_{n=1}^\infty \hbar\omega_n [ n (\omega_n)\delta(\omega+\omega_n) 
+ [n (\omega_n)+1]  \delta(\omega-\omega_n) ], 
\ee

where $n (\omega)=\mean{a^\dagger_n(0)a_n(0)}$ is the photon Bose-Einstein distribution 
with energy $\hbar\omega$ defined previously as $\mean{n}$ in eq.~\ref{Bose}. 

Taking the limit $\ell \rightarrow\infty$ and converting summation to integration through the replacement

\be 
 \sum_{n=1}^\infty f(\omega_n) \approx \frac{\ell}{v\pi} \int_0^\infty f(\omega) d\omega
\ee

yields 

\be 
S_{V}(\omega) = 2R \hbar\omega \{-n (-\omega) \Upsilon(-\omega) +[n (\omega)+1] \Upsilon(\omega) \}, 
\ee 

where $\Upsilon(\omega)=\int_0^\infty \delta(\omega-x) dx$ is the Heaviside step function.
Physically the negative $\omega$ term corresponds to energy absorption whereas in the
positive $\omega$ case,  $n(\omega)$ represents stimulated emission and +1 represents
spontaneous emission leading to $S_{V}(\omega)$ being asymmetric with respect 
to $\omega$ in contrast to the classical oscillator case (see Appendix A).

In the $\omega > 0$ case ($\Upsilon(\omega)$ term retained), the spectral density: 
\be 
S_{V}(\omega) = \frac{2R \hbar\omega}{1-e^{-{\hbar \omega / \kbt}} },
\ee 

reduces, in the classical limit~\cite{classical}  $k_{B}T \gg \hbar\omega$,
to Johnson-Nyquist noise result $S_{V}(\omega) = 2R k_{B}T$ (see note~\cite{factor2}).

In order to retrieve the quantum fluctuation-dissipation theorem\cite{Callen},
we take the symmetric part of $S_{V}(\omega)$ by adding positive and negative 
spectral contributions:

\be 
S_{V}(\omega) + S_{V}(-\omega) = 2R \hbar\omega \coth \left(\frac{\hbar\omega}{2k_{B}T} \right)
\ee

\section{Application to secure communications}
The main question in this section deals with the possible way to communicate
securely with a classical approach based upon acting on communication bits with 
controlled noise (shift-register generated pseudo-random bits) or through
Quantum Communications based on entanglement.

\subsection{Spread spectrum communications}
The principle of spread-spectrum communications such as DSSS used in Wi-Fi and 
cordless telephony is based on multiplying the message (made of 0's and 1's) 
by a sequence of pseudo-random bits. 
Pseudo-Random Binary Sequences (PRBS), the digital version of PN
sequences are produced in a controlled fashion
with a deterministic algorithm akin to pseudo-random numbers used in 
a Monte-Carlo algorithm or some other type of simulation~\cite{Recipes}.

The main goal of PRBS generation, is to draw 0 or 1 in an equally probable fashion
in order to have highly efficient crypting of the message (largest bandwidth or spreading). 
A particularly efficient method for producing PRBS is based on primitive polynomials modulo 2 
or Galois polynomials~\cite{Knuth} with the following arithmetic: \\
$0\oplus 0=0, 0\oplus 1=1, 1\oplus 0=1, 1\oplus 1=0$. \\
$\oplus$  is the usual symbol for modulo 2 arithmetic corresponding
to the logical XOR operation.

\begin{figure}[htbp]
  \centering
    \resizebox{70mm}{!}{\includegraphics[angle=0,clip=]{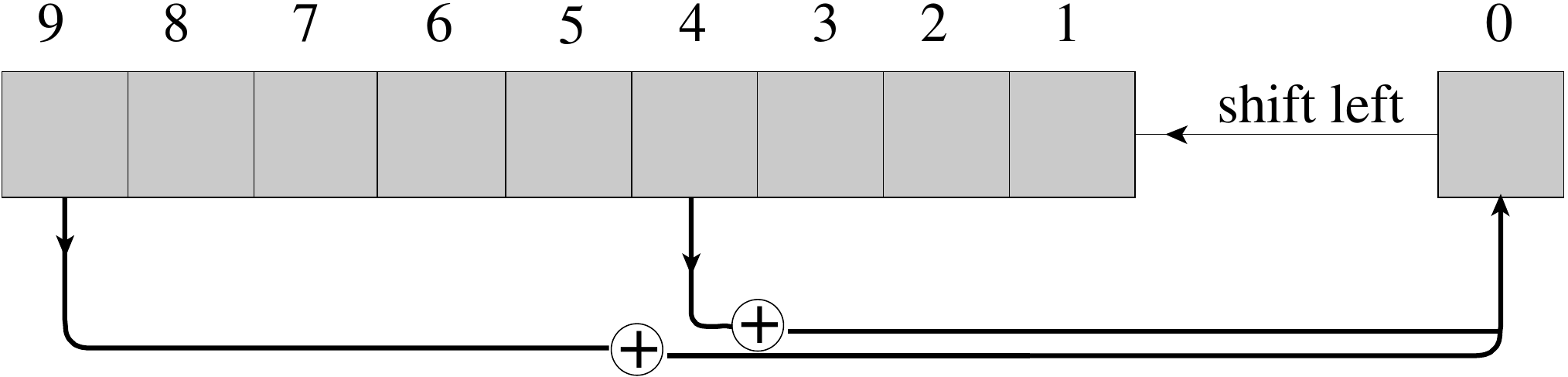}} 
\vspace*{-3mm}  
\caption{Shift register connections with feedback set up for PRBS~\cite{Recipes} generation 
on the basis of a (modulo 2) primitive polynomial given by $1+x^4+x^9$. It is of order 9 and  
tap connections (9,4,0) are shown.}
\label{XOR}
\end{figure} 

The coefficients of primitive polynomials modulo 2 are zero or one e.g. 
$x^4+x^3+1$, moreover they cannot be decomposed into a product of simpler modulo 2 polynomials.
An illustrative example is $x^2+1$ that cannot be decomposed into simpler
polynomials with real coefficients but can be decomposed into polynomials  with 
complex coefficients $x^2+1=(x+i)(x-i)$ with $i=\sqrt{-1}$. When this polynomial
is viewed as a Galois polynomial, it is not primitive since it can be decomposed into a product 
of simpler polynomials $x^2+1 \equiv x^2+2 x +1=(x+1)(x+1)$ since in modulo 2 arithmetic the term
$2 x$ is equivalent to 0 according to the above arithmetic rule ($1\oplus 1=0$).

The method for producing PRBS illustrated in fig.~\ref{XOR} requires only a single shift 
register $n$ bits long and a few XOR or mod 2 bit addition operations ($\oplus$ gates).

The terms that are allowed to be XOR summed together are indicated by shift register taps.
There is precisely one term for each nonzero coefficient in the primitive polynomial 
except the constant (zero bit) term. Table~\ref{Galois} contains a list of polynomials for $n \le 15$,
showing that for a primitive polynomial of degree $n$, the first and last term are 1.

\begin{table}[htbp]
\begin{tabular}{l|r}
\hline
Connection Nodes & Equivalent Polynomial \\
\hline
   (1,\hspace{0.3cm}0)  &  $1+x$ \\ 
    (2,\hspace{0.3cm}1,\hspace{0.3cm}0) &  $1+x+x^2$  \\  
    (3,\hspace{0.3cm}1,\hspace{0.3cm}0)  & $1+x+x^3$ \\
    (4,\hspace{0.3cm}1,\hspace{0.3cm}0)  & $1+x+x^4$\\ 
    (5,\hspace{0.3cm}2,\hspace{0.3cm}0)  & $1+x^2+x^5$\\    
    (6,\hspace{0.3cm}1,\hspace{0.3cm}0)  & $1+x+x^6$\\ 
    (7,\hspace{0.3cm}1,\hspace{0.3cm}0)  & $1+x+x^7$\\ 
    (8,\hspace{0.3cm}4,\hspace{0.3cm}3,\hspace{0.3cm} 2,\hspace{0.3cm} 0) & $1+x^2+x^3+x^4+x^8$ \\
    (9,\hspace{0.3cm}4,\hspace{0.3cm}0)  & $1+x^4+x^9$\\ 
    (10,\hspace{0.3cm}3,\hspace{0.3cm}0)  & $1+x^3+x^{10}$\\ 
    (11,\hspace{0.3cm}2,\hspace{0.3cm}0)  & $1+x^2+x^{11}$\\   
    (12,\hspace{0.3cm}6,\hspace{0.3cm}4,\hspace{0.3cm}1,\hspace{0.3cm} 0) & $1+x+x^4+x^6+x^{12}$ \\
    (13,\hspace{0.3cm}4,\hspace{0.3cm}3,\hspace{0.3cm}1,\hspace{0.3cm} 0) & $1+x+x^3+x^4+x^{13}$ \\ 
    (14,\hspace{0.3cm}5,\hspace{0.3cm}3,\hspace{0.3cm}1,\hspace{0.3cm} 0)  & $1+x+x^3+x^5+x^{14}$\\ 
    (15,\hspace{0.3cm}1,\hspace{0.3cm}0) &  $1+x+x^{15}$\\  
\hline
\end{tabular}
\caption{List of the first 15 Galois polynomials.}
\label{Galois}
\end{table}

A Maximum-Length Sequence (MLS) $x[n]$ is a balanced sequence made from 
equally probable symbols with values +1 and -1 such that the MLS averages to zero. 
Choosing $x[n] = (-1)^{a[n]}$ with $a[n] =$ 0 or 1 originating from PRBS yields the 
desired values $x[n]$ = +1 or -1 with +1/-1 equally probable.
The PRBS sequence $a[n]$ is produced with a shift register XOR operation 
as discussed previously and illustrated in fig.~\ref{XOR}.
The MLS has many attractive features in addition to the balanced character: 
its standard deviation and peak values are both equal to 1 making its crest factor 
(peak/standard deviation) equal to 1, the lowest value it can get~\cite{Recipes}. 
That is why MLS has noise-immune property~\cite{Recipes} required in communication
electronics. MLS are used not only in secure communications but also in synchronization 
of digital sequences. 

\begin{figure}[htbp]
  \centering
    \resizebox{60mm}{!}{\includegraphics[angle=0,clip=]{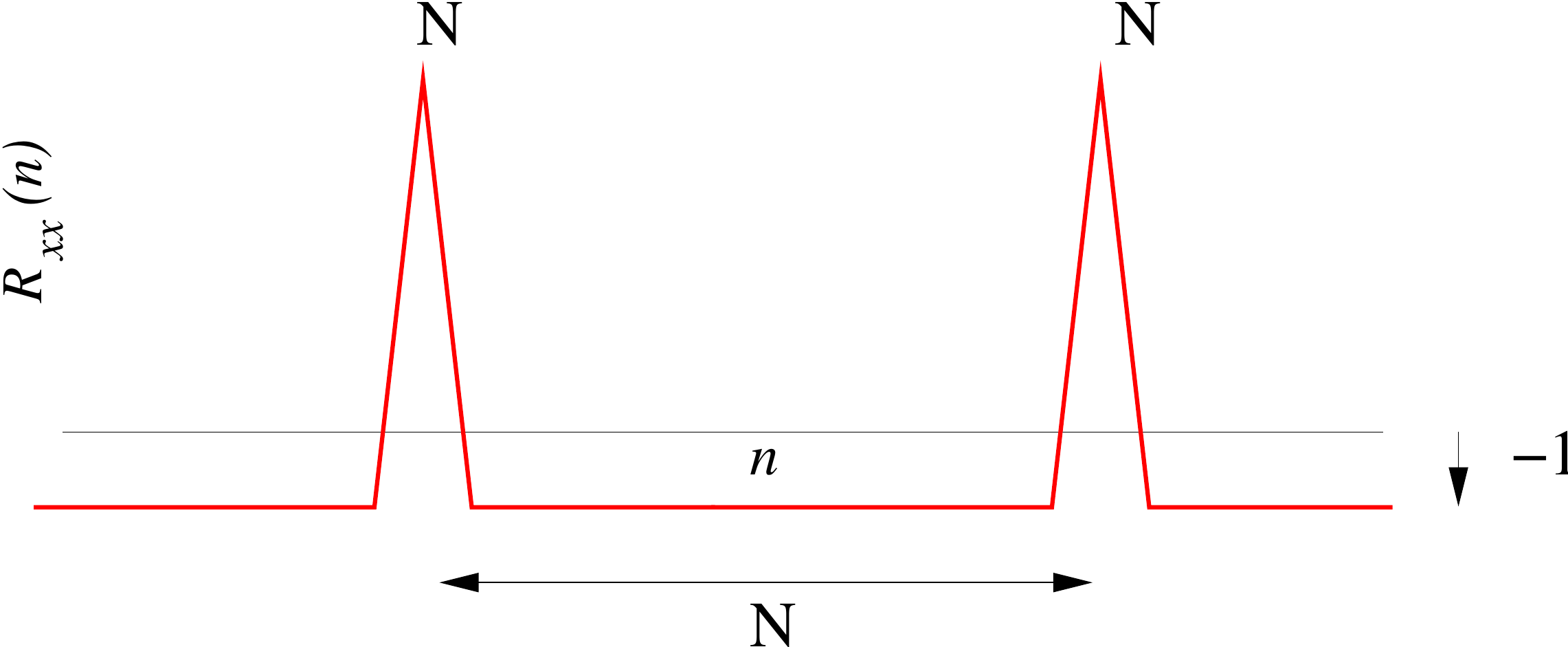}} 
\vspace*{-3mm}  
\caption{Property of MLS auto-correlation $R_{xx}[n]$ showing peaks that enables decoding
the message and displaying maximum length $N=2^{n_c} -1$ with $n_c$ the number of coding bits.}
\label{MLS}
\end{figure} 

A message $x(t)$ transmitted through a linear time-invariant medium is    
convoluted with the channel impulse response $h(t)$ resulting in an output message:

\be
y(t)=h(t)*x(t)=\int_{-\infty}^{\infty}h(t-t') x(t') dt'
\label{reception}
\ee

The decoding process of the message is based on a correlation operation
based on the $x[n]$ auto-correlation given by: 
                         
\bea
R_{xx}[n] &=& \frac{1}{N-1}  \sum_{i=0}^{N-2}{ x[i] x[n+i] }   \nonumber \\
         & =& \frac{1}{N-1}  \sum_{i=0}^{N-2}{ (-1)^{(a[i] \oplus a[n+i])} }                         
\eea

with $N=2^{n_c} -1$ where $n_c$ is the number of coding bits or MLS order. $N$ is the period or the
length of the MLS.

\begin{figure}[htbp]
  \centering
    \resizebox{60mm}{!}{\includegraphics[angle=0,clip=]{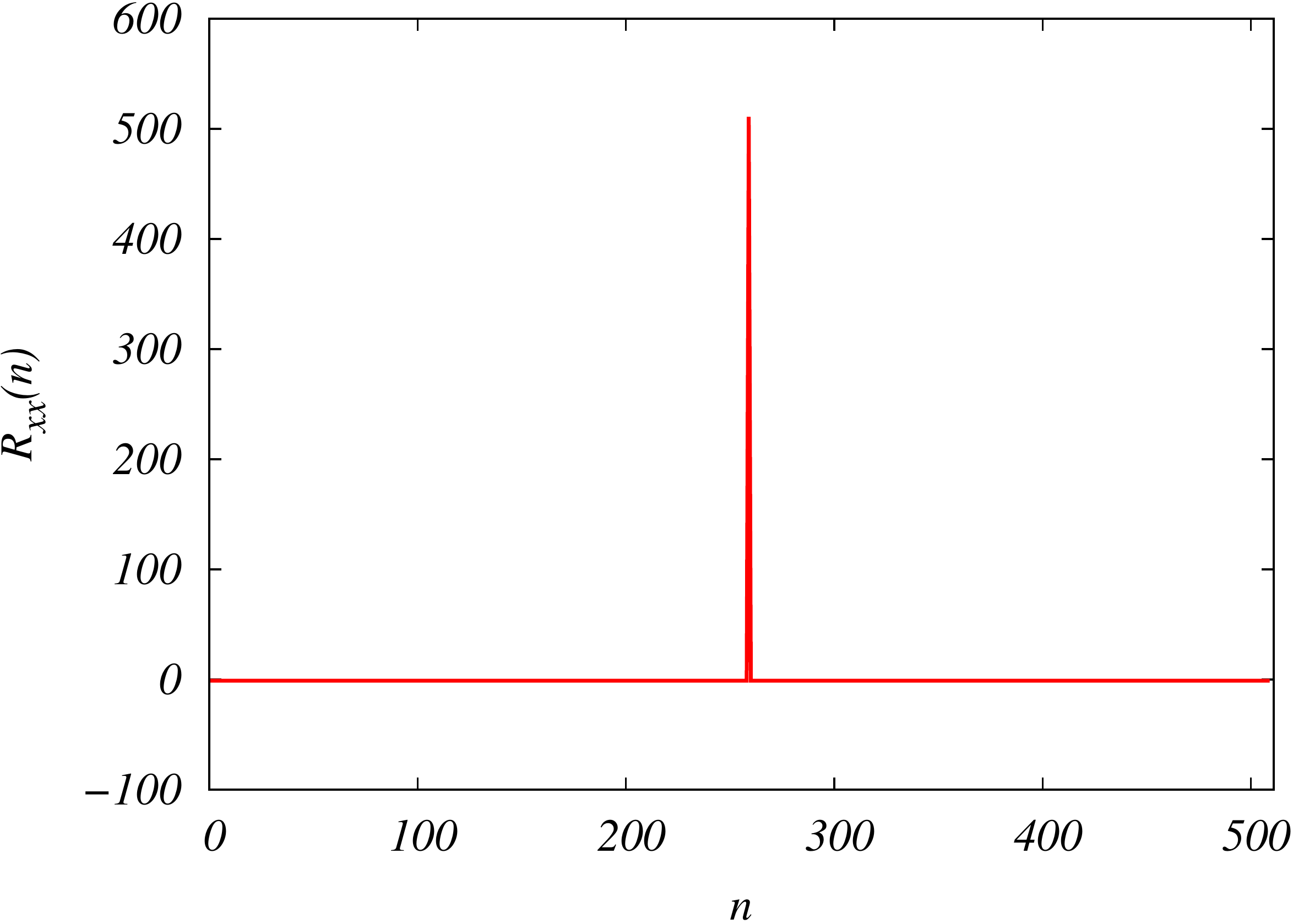}} 
\vspace*{-3mm}  
\caption{Auto-correlation $R_{xx}[n]$ of order 9 MLS displaying the peak value of 2$^9$-1=511 
over the period interval [0-511].}
\label{Auto}
\end{figure}

As an example, the auto-correlation  $R_{xx}[n]$ of order $n_c=9$ shown in fig.~\ref{Auto} displays a 
$\delta$ function-like behaviour required for message decoding or synchronization (shown in fig.~\ref{MLS}).

Another application  of the MLS  is the determination of the impulse response $h(t)$ of any communication channel 
by sending through the channel a PRBS signal $x(t)$ whose auto-correlation is a delta function that will be used
to identify $h(t)$ at the receiver (see eq.~\ref{reception}) since:

\be
h(t)=\int_{-\infty}^{\infty}h(t-t') \delta(t') dt'
\label{identification}
\ee


The impulse response determined with MLS is known to be immune to distortion. This is why 
despite the fact, many other methods~\cite{Stan} exist to measure it with various success, 
MLS is still preferred when distortion is an issue. 


\subsection{Quantum communication (QC)}
Quantum Mechanics provides several important ingredients to information communication
not present in its classical counterpart~\cite{Ekert}.
Firstly the information itself sent across a communication channel can be either 
classical or quantum. The same applies to the channel that might be classical or quantum.
Information transmission is measured with input-output correlations performed 
across the channel that can also be classical or quantum, the signature of entangled states. 
Copying a bit in classical communication is a trivial voltage replication operation
whereas in QC the no-cloning theorem~\cite{Scarani} forbids copying quantum information
without leaving a trace. Crypting information can be made with classical keys (as in PRBS) or
quantum keys. The generation of random numbers through quantum means
(QRNG) are superior to PRBS despite their many interesting properties.

Quantum networks across which quantum information is carried is also different from its
classical counterpart and finally classical noise as well as quantum noise should be
properly described in order to evaluate information error rates.

We describe every element of quantum communication below.

\subsubsection{Quantum unit of information: the qubit}

\begin{figure}[htbp]
  \centering
    \resizebox{60mm}{!}{\includegraphics[angle=0,clip=]{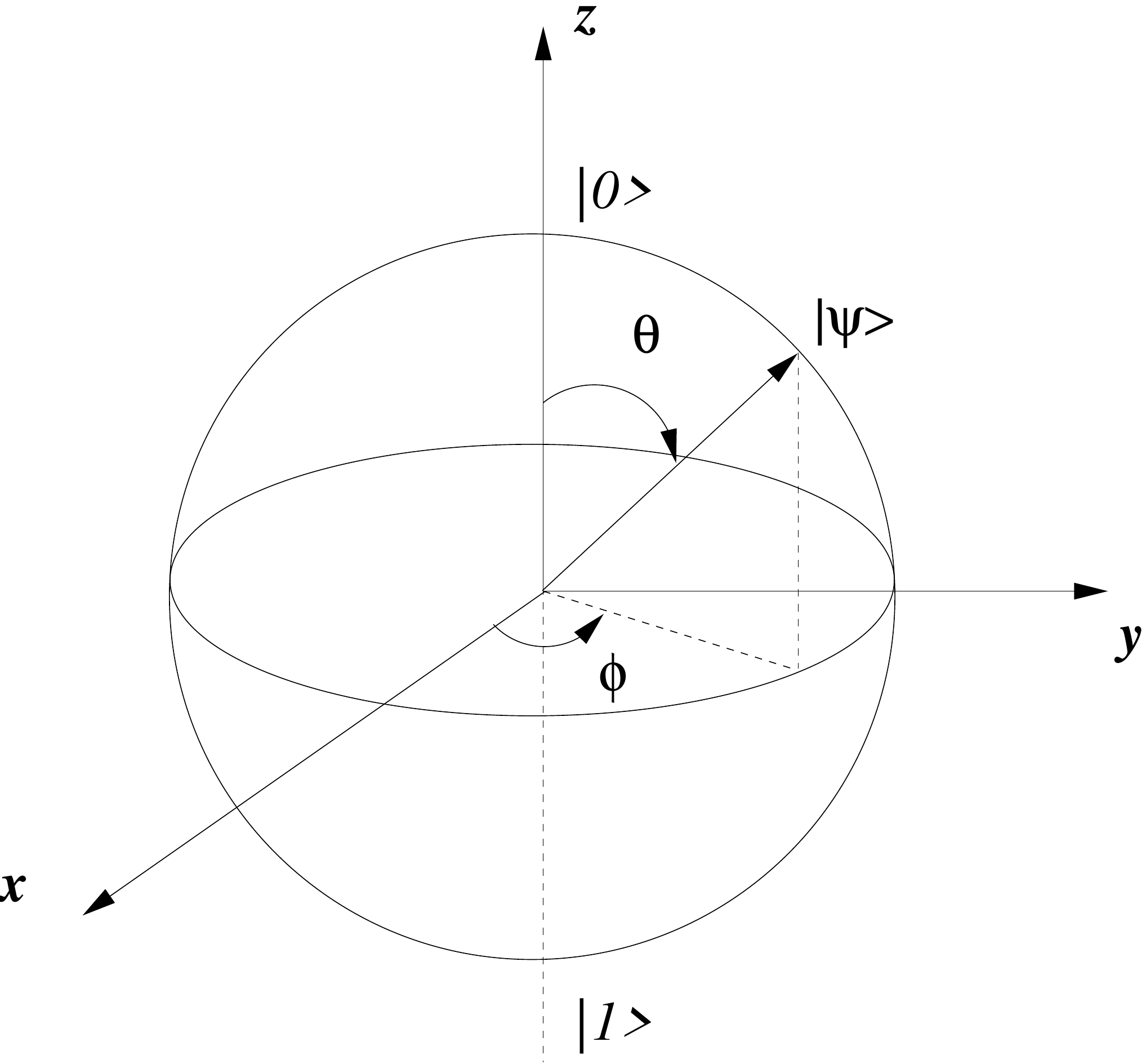}} 
\vspace*{-3mm}  
\caption{Bloch sphere (Poincar\'e sphere for photons) representing the possible 
values of a quantum information unit in 2D Hilbert space or qubit shown as the
quantum wavefunction $\ket{\psi(\theta,\phi)}$. 
The classical bits (0,1) are the poles of the sphere. A qubit is
any two-component wavefunction given by $(\alpha \ket{0} + \beta \ket{1})$ 
with complex coefficients $\alpha,\beta$. A pure state exists over the Bloch sphere with
$|\alpha|^2 +|\beta|^2 =1$ whereas a mixed state lies inside the sphere with
$|\alpha|^2 +|\beta|^2 <1$. A classical
analog~\cite{continuous} state lies anywhere on the vertical axis linking the poles.}
\label{bloch}
\end{figure}

The discrete~\cite{continuous} unit quantum information in 2D Hilbert space is the qubit, 
the two-state quantum counterpart of the classical bit (see fig.~\ref{bloch}). 

It is represented by a 
two-component wavefunction (or spinor~\cite{spinor}) $\ket{\psi(\theta,\phi)}$.

Computationally, a qubit is representable with 
128 classical bits considering that it is made of two complex numbers that 
are equivalent themselves to four 32 bit (single precision) float numbers.

In the case of photons, quantum states $\ket{0}$ and  $\ket{1}$ are equivalent 
to orthogonal polarization states (see fig.~\ref{bloch}).
  
The photon is the logical choice as the basic information carrier in quantum
communications proceeding between nodes that make quantum networks. 
Information can be encoded in photon polarization, orbital momentum, spatial mode 
or time and any manipulation targeting processing
or information transfer can be made with optical operations, 
such as using birefringent waveplates to encode polarization...

On the other hand, atoms are the natural choice to make quantum memories since some of 
their electronic states can retain quantum information for a very long time.

Quantum networks convey quantum information with nodes that allow for its 
reversible exchange. The latter may be done with two coupled single-atom nodes 
that communicate via coherent exchange of single photons. In comparison, classical
fiber-optic networks use pulses containing typically 10$^7$ photons each.

In order to prevent change in information or even its loss, it is necessary 
to have tight control over all quantum network components. Considering the 
smallest memory for quantum information as a single atom with single photons 
as message carriers, efficient information transfer between an atom and a 
photon requires strong interaction between the two components not achievable 
with atoms in free space but in special optical cavities.

A low-loss cavity made with a set of strongly reflective mirrors  
alters the distribution of modes with which the atom interacts modifying 
the density of vacuum fluctuations that it experiences at a given frequency
enhancing or reducing atomic radiative properties. As a consequence, 
spontaneous emission from the atom excited state being a 
major source of decoherence can be inhibited in a cavity. 

A low-loss optical cavity possesses a high quality factor ($Q > 10^3$) allowing
a photon entering the cavity to be reflected between mirrors making the cavity several 
thousand times per second strongly enhancing its coupling with the atom leading
to its absorption by the atom in a highly efficient coherent fashion. 

On the other hand, photon emission by an atom inside a cavity is highly directional
and can be sent to other network nodes in a precisely controlled fashion. 


Controlling qubit states means that an operator is required to allow switching from one qubit state to another. 
A rotation matrix $\hat{R}_z(\theta,\phi)$ represents such an operator in the $\ket{0},\ket{1}$
basis:

\begin{equation}
   \hat{R}_z(\theta,\phi)= \left [
   \begin{array}{cc}
   \cos\frac{\theta}{2}&-ie^{i\phi}\sin\frac{\theta}{2}\\
   -ie^{-i\phi}\sin\frac{\theta}{2}&\cos\frac{\theta}{2}
   \end{array}\right ] \label{rotation matrix}
\end{equation} 

Applying $ \hat{R}_z(\theta,\phi)$ on the state $\ket{0}$ allows us 
to produce an arbitrary state $(\theta \ne 0,\phi \ne 0)$  on the Bloch sphere: 
$\ket{\psi(\theta,\phi)}=  \hat{R}_z(\theta,\phi) \ket{0}= 
\cos\frac{\theta}{2}\ket{0} -ie^{-i\phi}\sin\frac{\theta}{2}\ket{1} \equiv (\alpha \ket{0} + \beta \ket{1})$. 
This applies only for pure states that lie on the
Bloch sphere surface ($|\alpha|^2 +|\beta|^2 =1$). In the case of mixed states 
($|\alpha|^2 +|\beta|^2 <1$) we need additional operators that alter also the 
wavefunction modulus. Experimentally, photon polarization can be rotated
with a half-wavelength plate (called also a Hadamard gate), moreover it can be
separated into individual components with a polarizing beam-splitter 
(see ref.~\cite{Obrien}).

\subsubsection{Entanglement}

Quantum networks possess special features that are not found in their classical counterparts. 
This is due to the intrinsic nature of the information processed: 
while a classical bit is either 0 or 1 (see fig.~\ref{bloch}), a qubit 
(quantum bit wavefunction) can take both values at the same time due to coherent 
superposition inherent to Quantum Mechanics linearity.
 
Quantum Mechanics embodies the notion of entanglement 
detected with violation of Bell inequalities~\cite{Bell} and
that brings a paradigm shift into information processing.

Given two qubits $\ket{Q_1}=~\frac{1}{\sqrt{2}}(c_0 \ket{0} + c_1 \ket{1})$ 
and $\ket{Q_2}~=~\frac{1}{\sqrt{2}}(d_0 \ket{0} + d_1 \ket{1})$, 
it is possible to build a state $\ket{Q_1,Q_2}=\ket{Q_1}\otimes \ket{Q_2}$ such that:
\bea
\ket{Q_1,Q_2}=\frac{1}{\sqrt{4}}(c_0 \ket{0} + c_1 \ket{1})\otimes (d_0 \ket{0} + d_1 \ket{1}) \nonumber \\
 =\frac{1}{\sqrt{4}}(c_0 d_0 \ket{0,0}+c_0 d_1 \ket{0,1}+c_1 d_0 \ket{1,0}+c_1 d_1 \ket{1,1}) \nonumber \\
\eea

Quantum Mechanics, however, allows for building other states such as:
$\ket{Q}=\frac{1}{\sqrt{2}}(c_0 d_1 \ket{0,1}+c_1 d_0 \ket{1,0})$
which are not decomposable into products of constituent states. 

These states are called entangled and can be mapped onto the polarization of single photons 
which can be transferred through an optical fiber between two nodes consisting respectively 
of atoms in state $\ket{A}$ and state $\ket{B}$. 

Quantum mechanical entanglement ought to be achieved between the two nodes in order to
have successful QC maintained over the coherence time preserving
integrity of quantum information transfer.

In order to achieve entanglement between two remote network nodes, polarization of the 
single photon emitted by atom in state $\ket{A}$ is entangled with the atomic quantum state. 

Once the photon gets absorbed, the entanglement is transferred onto atom in state $\ket{B}$
and reversible exchange of quantum information is performed between the two nodes. 


Experimental production of entanglement can be made between two particles 
(bipartite) or between several particles (multipartite) and for each number of particle 
(two, three, four ...) case or  particle type (atoms, electrons, photons...) 
several experimental procedures readily exist. It is not limited to microscopic
particles since it can be induced by a light pulse between two macroscopic 
objects~\cite{Julsgaard} consisting each of a gas containing 
about 10$^{12}$ Cesium atoms.

Entanglement can occur when particles interact and kept in contact or 
when they emerge from a common ancestor as in the EPR~\cite{EPR} case where 
a spinless particle decays into two particles carrying opposite spins... 
Another example is the case of a photon interacting with a non-linear crystal. 
It can be destroyed and replaced with a lower energy entangled photon pair 
(the process is called SPDC~\cite{Kwiat} or Spontaneous Parametric Down-Conversion).

Heralded~\cite{Monroe} entanglement may occur between non-interacting remote particles
(Yb ions held in two ion traps, 1 meter apart) not possessing a common ancestor, 
however the entanglement probability $p_E$ is very low ($p_E \approx 10^{-9}$) since
entanglement results from the interaction of decay photons emitted by each ion after 
their excitation by picosecond laser pulses. Thus $p_E$ needs to be increased substantially 
in order to make it applicable to mass QC. 


\subsubsection{Quantum Random Number Generation}
Classical Random Number Generators (RNG) are based on Uniform
RNG and the standard statistical quality tests target 
the uniformity~\cite{Recipes} of the numbers generated.
Quantum Mechanics introduce a predictability test to further improve
quality of RNG.

This means that even if the RN is perfectly uniformly distributed, 
it may contain hidden deterministic information and is therefore 
prone to be predictable. For instance, PN and PRBS generate uniformly 
distributed numbers but since they are produced with a deterministic 
algorithm, an eavesdropper might, by drawing values and performing
statistical analysis~\cite{Recipes}, be able to make an educated guess 
and access the cipher password, key...

Thus, statistical uniformity tests are necessary but not sufficient to guarantee that any given 
RNG is not prone to attack and guess by an intruder. Quantum RNG (QRNG) offers "true
RN" generation that is very difficult to predict. Using a special program called 
"randomness extractor"~\cite{Colbeck} one might eliminate all bit strings originating from 
an implicit deterministic algorithm and keep only truly random bit strings. For this reason,
the method is also called, amplification of weak randomness~\cite{Colbeck}.

Randomness extraction procedure exploits entropy hierarchy (see Appendix D) that
attributes a number of bits depending on the entropy estimation used. 
R\'enyi min-entropy is very efficient computationally wise
and a string of perfectly random bits has unit min-entropy per bit as derived in Appendix D.

Starting from $l$ input bits $X_i$ of low-entropy per bit ($s < 1$), the extractor 
computes a number $k < l $ of higher-entropy ($s' \approx 1$) output bits $Y_j$ with
a linear transformation via multiplication by a matrix $m$:
\be
Y_j = \sum^l_{i=1}m_{ji}X_i, \hspace{1cm} j=1...k
\label{extractor}
\ee

$m$ is built from $l\times k$ random bits that can be generated with Galois
polynomials and all arithmetic operations are done modulo 2 with AND and XOR logic.

This "Whitening" procedure can be viewed as the quantum counterpart of the Maximum Entropy
Method that is widely used in Image Processing for deblurring images~\cite{Recipes}.

As a direct application of this concept, Sanguinetti \etal~\cite{Sanguinetti} used 
Smartphone cameras to produce Quantum Random Numbers. 
After uniform illumination of the camera image sensor 
by a LED and estimation of the number of photons generated per pixel, a randomness 
extractor algorithm such the above (eq.~\ref{extractor}) is used to 
compute truly random numbers.

For $X_i$ input bits with low entropy per bit ($s < 1$), the probability that 
the output $Y_j$ deviates from a perfectly random bit string (with high entropy
per bit $s' \approx 1$) is bounded~\cite{Cover} by:
\begin{equation}
\epsilon=2^{-(l s - k s')/2},
\end{equation}

Picking a CCD image sensor with 16 bits per pixel (detection capability) 
and a photon flux producing $2\times 10^4$ electrons per 
pixel gives $R_\infty=8.469$ bits/pixel (from eq.~\ref{Renyi}) 
yielding a min-entropy per bit $s=0.529$  (in comparison, Shannon Entropy 
is  9.191 bits/pixel or 0.574 per bit). Selecting input $l=2000$, 
output $k=400$ and $s'=1$, we get $\epsilon =2.57 \times 10^{-197}$.

As a result, an eavesdropper would have to generate 
an extremely large~\cite{Cover} amount of random numbers (about $1.97 \times 10^{99}$)
before noticing any departure from a perfectly random sequence, indicating
the superior performance of QRNG with respect to any classical RNG.


\subsubsection{Quantum Keys}
Classical cryptography is based on two types of keys that are used to
encode and decode messages: secret or symmetric keys and public or asymmetric keys.
Symmetric keys are same for encoding and decoding messages whereas in 
public cryptography systems, one needs a public key and a private key.
In the PGP (Pretty Good Privacy) secure mailing system over the Internet,
the sender encodes the message with receiver public key and the receiver
decodes the message with his private key. 
In quantum cryptography, the simplest example of secret key sharing among
sender and receiver (Alice and Bob) in QKD is the BB84~\cite{Scarani} protocol. 
Alice and Bob communicate through two channels: one quantum to send
polarized single photons and one classical to send ordinary messages.
Alice selects two bases in 2D Hilbert space consisting each of two
orthogonal states: $\bigoplus$ basis with $(0,\pi/2)$
linearly polarized photons,   
and $\bigotimes$ basis with $(\pi/4, -\pi/4)$ linearly polarized photons.

Four symbols: $\ket{\rightarrow}, \ket{\uparrow},\ket{\nearrow},\ket{\searrow}$
representing polarized single photons are used to transmit quantum data with
$\ket{\nearrow}=\frac{1}{\sqrt{2}}(\ket{\rightarrow}+ \ket{\uparrow})$
and $\ket{\searrow}=\frac{1}{\sqrt{2}}(\ket{\rightarrow}- \ket{\uparrow})$.

In the $(basis,data)$ representation, the symbols are given by 
$\ket{\rightarrow}=(\bigoplus,0)$, $\ket{\uparrow}=(\bigoplus,1)$ in the $\bigoplus$
basis whereas $\ket{\searrow}=(\bigotimes,0)$, $\ket{\nearrow}=(\bigotimes,1)$ in the
$\bigotimes$ basis.

A message transmitted by Alice to Bob over the Quantum channel is a stream of 
symbols selected randomly among the four described above. 

Bob performs polarization measurements over the received symbols selecting 
randomly bases $\bigoplus$ or $\bigotimes$.

Afterwards Bob and Alice exchange via the classical channel their mutual
choice of bases without revealing the measurement results.

In the ideal case (no transmission errors, no eavesdropping) 
Alice and Bob should discard results pertaining to 
measurements done in different bases (or when Bob failed to detect 
any photon).  This process is called "key sifting" after which the raw key is determined.

After key sifting, another process called key distillation~\cite{Scarani} must be performed.
This process entails three steps~\cite{Scarani}: error correction, privacy amplification and
authentication in order to reveal classical or quantum errors of transmission,
detect eavesdropping (with the no-cloning theorem~\cite{Scarani}) and act against it. 

Ignoring, for simplicity, key distillation, the raw key size is typically 
about one quarter of the data sent since both Alice and Bob are selecting 
their bases at random (total probability is roughly  
$\frac{1}{2} \times \frac{1}{2}=\frac{1}{4}$).


A random number generator (RNG) or rather
a random bit generator can be used to select $\bigoplus$ or $\bigotimes$ bases. 
Using PRBS or, even better, QRNG to select measurement bases, we infer that
by comparison with the classical FHSS crypting method, Quantum Mechanics 
provides extra flexibility through basis selection. 
Such option is simply not available in classical communication.


On the negative side, there are several problems that may come up with the 
BB84 scheme. One major obstacle is that
presently, it is difficult, on a large scale level, to produce single photons.  
One approximate method for doing this, is to 
use attenuated laser pulses containing several photons that might be intercepted
in the quantum channel by an eavesdropper with a PNS (Photon Number Splitting) attack. 

Quantum Communications can be made more secure when QKD is implemented with
entanglement~\cite{Scarani} providing a secure way to distribute secret keys between remote 
users such that when some eavesdropper is detected, the transmission is halted and the data discarded.


The BBM92~\cite{Scarani} scheme is an entanglement based version of the BB84 protocol. 
Polarization entangled photon pairs (called EPR pairs or Bell states) are sequentially 
generated with one photon polarization measured by Alice and the other 
measured by Bob. EPR pairs are produced after emerging from SPDC~\cite{EPR} by using 
a birefringent phase shifter or slightly rotating the non-linear crystal 
itself since the state produced by SPDC is:

\be
\ket{\psi}=\frac{1}{\sqrt{2}}(\ket{\rightarrow \uparrow} + e^{i \varphi} \ket{\uparrow \rightarrow})
\ee

Thus it suffices to modify $\varphi$ to 0 or $\pi$ or place a quarter wave-plate 
giving a 90\deg shift in one photon path to generate all 
Bell states~\cite{Kwiat}.
These states are polarization entangled~\cite{Scarani} photons :

\be
\ket{\psi^\pm}=\frac{1}{\sqrt{2}}(\ket{\rightarrow \uparrow} \pm \ket{\uparrow \rightarrow}),
\ket{\phi^\pm}=\frac{1}{\sqrt{2}}(\ket{\rightarrow \rightarrow} \pm \ket{\uparrow \uparrow})
\ee

The set forms a complete orthonormal basis in 4D Hilbert space for all polarization states
of a two-photon system.

Alice and Bob  choose randomly one of the two bases $\bigoplus$ or $\bigotimes$
to perform photon polarization measurement.

Afterwards Alice and Bob communicate over the classical channel 
which basis they used for each photon successfully received by Bob.

The raw key is obtained by retaining the results obtained when the bases used are same.
Neither RNG nor QRNG are used in this case since randomness is inherent 
to the EPR pair polarization measurement~\cite{Scarani}.
Moreover, no Bell inequality tests are needed since all measurements 
must be perfectly correlated or anti-correlated.

For instance in the $\ket{\psi^+}$ state, if one photon is measured to be in the 
$\ket{\rightarrow}$ state, the other must be in the $\ket{\uparrow}$ since 
the probabilities of measuring ${\rightarrow \rightarrow}$ or ${\uparrow \uparrow}$ are given by 
$|\bra{\rightarrow \rightarrow}\ket{\psi^+}|^2=|\bra{\uparrow \uparrow}\ket{\psi^+}|^2=0$, 
whereas the probabilities of measuring ${\rightarrow \uparrow}$ and ${\uparrow \rightarrow}$ are 
$|\bra{\rightarrow \uparrow}\ket{\psi^+}|^2=|\bra{\uparrow \rightarrow}\ket{\psi^+}|^2=\frac{1}{2}$.
This is termed perfect anti-correlation.

When the polarization measurements are performed in the $\bigotimes$ basis, we get
rather, perfect correlation. That means 
the probabilities of measuring ${\nearrow \searrow}$ or ${\searrow \nearrow}$ are 
$|\bra{\nearrow \searrow}\ket{\psi^+}|^2=|\bra{\searrow \nearrow}\ket{\psi^+}|^2=0$,
whereas the probabilities of measuring ${\nearrow \nearrow}$ or ${\searrow \searrow}$ are given by 
$|\bra{\nearrow \nearrow}\ket{\psi^+}|^2=|\bra{\searrow \searrow}\ket{\psi^+}|^2=\frac{1}{2}$. 

Note that if we rather consider the  $\ket{\psi^-}$ state, we get perfect 
anti-correlation in both bases $\bigoplus$ and $\bigotimes$.


\subsubsection{Quantum Networks}

In classical communications, channel transfer function, the Fourier transform
of its impulse response $h(t)$ is a function of
frequency and distance. Channel bandwidth and signal attenuation are functions
of distance. When a pulse (representing a communication symbol made of several
bits depending on the modulation method used) is sent through an optical fiber, 
it undergoes broadening leading to inter-symbol interference, attenuation
leading to signal loss and alteration due to noise. 
Thus it is required to evaluate the largest distance
that could be covered at the end of which a repeater is placed in order to filter
out noise and restore pulse shape to its original form.

In QKD, Alice and Bob should be able to determine efficiently their shared secret key 
as a function of distance $L$ separating them. Since, the secure key is determined after
sifting and distillation, secure key rate is expressed in bps (bits per symbol) given
that Alice sends symbols to Bob to sift and distill with the remaining bits making the secret key.

The simplest phenomenological way to estimate secure key rate versus distance $K(L)$ is to consider 
a point-to-point scenario with $K(L) \propto [A(L)]^n$  where $A(L)= 10^{-\alpha_0 L/10}$
is signal attenuation versus distance. $\alpha_0$ is the attenuation 
coefficient per fiber length and $n=1,2...$. $\alpha_0$ depends strongly on the wavelength $\lambda$
used to transmit information through the fiber. For the standard Telecom wavelength~\cite{Carlson}
$\lambda=1.55 \mu$m, $\alpha_0$=0.2 dB/km.

The optimal distance~\cite{Scarani} $L_{opt}$ is determined by the maximum of the objective
function $L K(L)$. Taking  the derivative and solving, we get 
${L_{opt}=\frac{10}{n \alpha_0 \ln(10)}}$.

This yields $L_{opt}=$21.7 kms for $n=1$, $L_{opt}=$10.86 kms for $n=2$ and 
$L_{opt}=$5.43 kms for $n=4$. 

Errors produced by noise, interference and damping are represented by a $BER$
(Bit Error Rate), the ratio of wrong bits over total number of transmitted bits.
$BER$ versus distance is an important indicator of communication quality as much as
communication speed is represented by bit rate versus distance. 

\begin{figure}[htbp]
  \centering
    \resizebox{80mm}{!}{\includegraphics[angle=0,clip=]{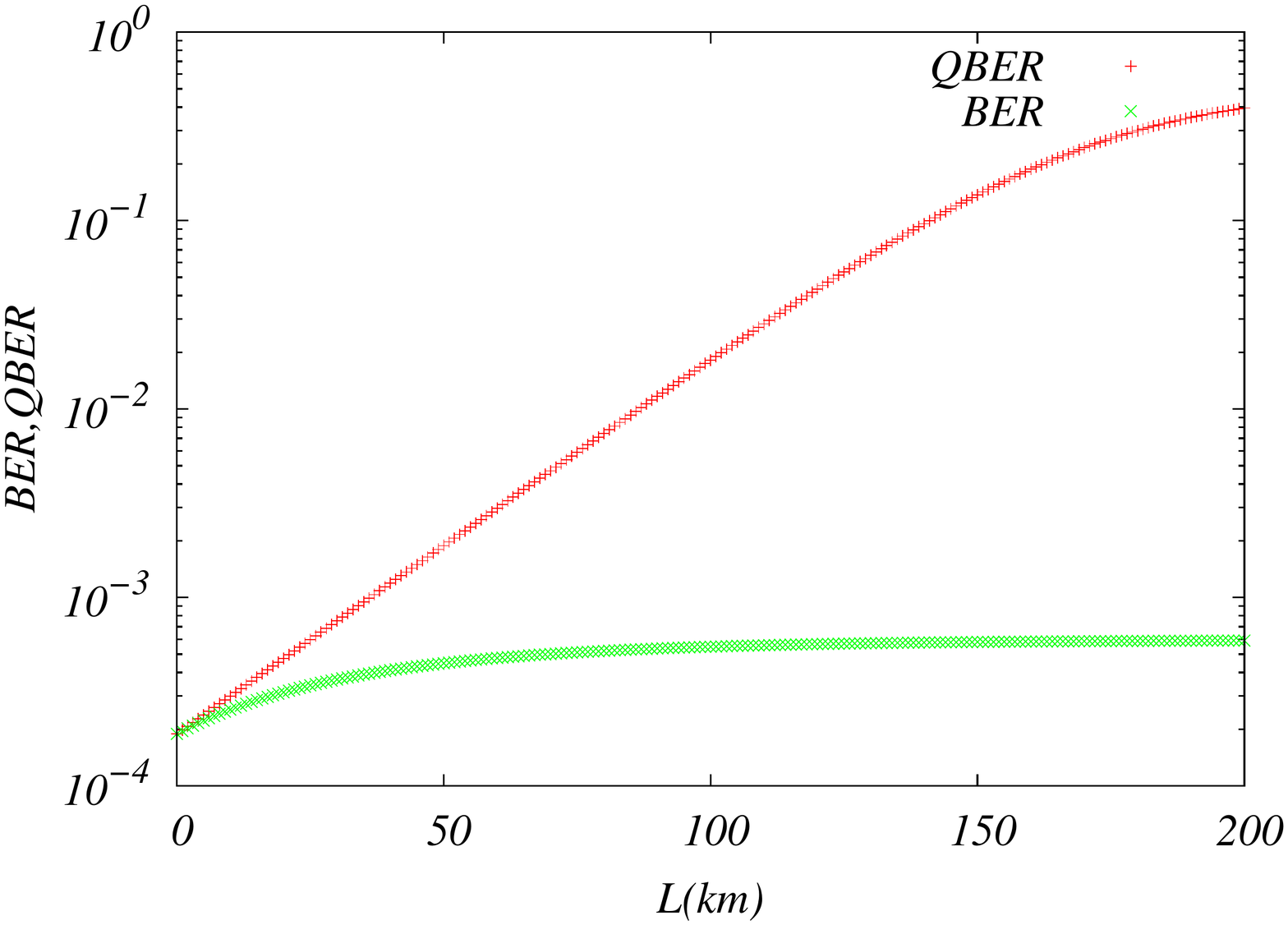}} 
\vspace*{-3mm}  
\caption{(Color on-line) Classical $BER$ and quantum $QBER$  versus distance
$L$ along an optical fiber using the BB84 protocol considering they start from
the same value at $L=0$.}
\label{QBER}
\end{figure}

In the quantum case, the $QBER$ (Quantum $BER$) $Q_e(L)$ versus distance 
is the quantity of interest. Regarding the BB84 protocol case, a simple model~\cite{GYS} 
delivers the expression:

\be
Q_e(L)=\frac{P_e}{ A(L) \mu  \eta_{Bob}  +2P_e}
\ee

where $P_e$ is the probability of error per clock cycle (measured to be
8.5 $\times 10^{-7}$). $\mu=0.1$ is the  average photon flux used by Alice
to transmit symbols and $\eta_{Bob}=0.045$ is Bob apparatus detection efficiency. 
The results are displayed versus distance $L$ in fig.\ref{QBER}.

Fig.~\ref{QBER} shows that the $QBER$ increases faster and takes larger values 
than the optical fiber classical $BER$. For many digital lightwave systems
using ON-OFF modulation~\cite{Carlson} (1 for light pulse, 0 for no pulse), the classical $BER$ 
is typically about $10^{-9}$ and may reach values in the $[10^{-16}-10^{-15}]$ range.

\begin{figure}[htbp]
  \centering
    \resizebox{80mm}{!}{\includegraphics[angle=0,clip=]{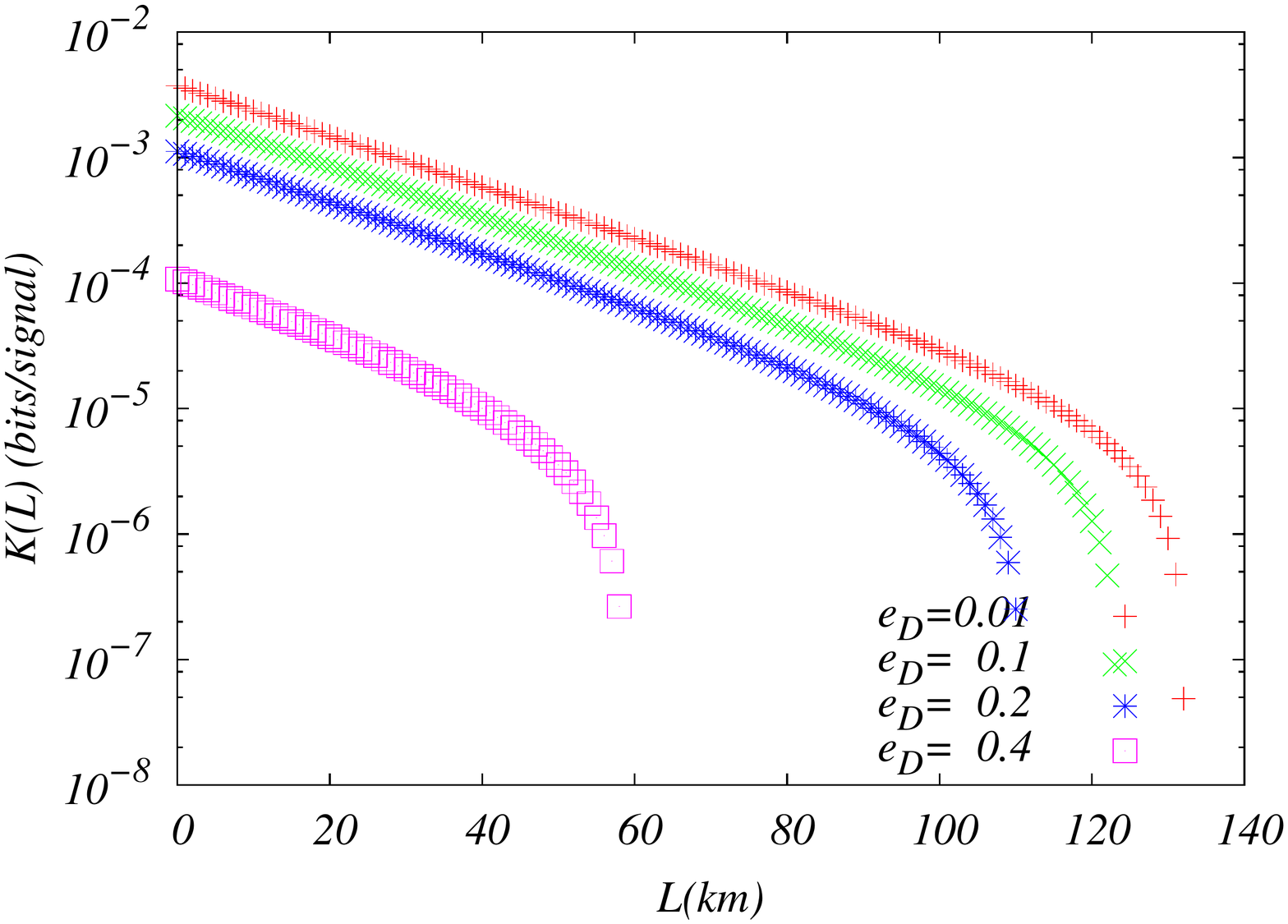}}  
\vspace*{-3mm} 
\caption{(Color on-line) Key rate $K(L)$ in bps versus distance $L$ using the same parameters 
as in Ref.~\cite{GLLP} for several detector error rates $e_D$=0.01, 0.1, 0.2 and 0.4.}
\label{Rate}
\end{figure}

Moving on to estimate the secure key generation rate in bits per symbol (bps) emitted 
by Alice, a simple model for the BB84 protocol~\cite{GLLP} gives:

\be
K(L)=G_\mu \{ -h_2(Q_e(L)) + \Omega [1-h_2(e_1)]  \}
\ee

where $G_\mu$ is the gain for an average photon flux $\mu$. 
$\Omega$ is the fraction of events detected by Bob and produced
by single-photon signals emitted by Alice. $e_1$ is the corresponding $QBER$ and
$h_2$ is the binary Shannon entropy~\cite{Carlson} given by 
$h_2(x)=-x\log_2(x)-(1-x)\log_2(1-x)$. Using the same parameters as in Ref.~\cite{GYS} and 
bounds for $\Omega$ and $e_1$ estimated in Ref.~\cite{GLLP}, we are able to plot 
the secure key rate versus distance for several
values of the detector error rate $e_D$ as displayed in fig.~\ref{Rate}.

Fig.~\ref{Rate} shows that the key rate is small and given that security increases 
with key length, a major improvement with respect to this simple approach 
should be undertaken in order to increase substantially the bps rate.

Recently, a joint team from Cambridge Science Park and University of Cambridge~\cite{Comandar} 
succeeded to increase substantially the secure key rate using detectors operating at room temperature.
The secure key rates obtained were between 1.79 Mbit/s and 1.2 kbit/s for 
fiber lengths between 40 km and 100 km, respectively.



Regarding network building developments, the first elementary quantum network based on interfaces 
between single atoms and photons located at two network nodes installed in two distant 
laboratories connected by an optical fiber link was made in 2012 by a team of 
scientists~\cite{Ritter} at the Garching MPQ.

Using the above procedures, Ritter \etal~\cite{Ritter} were able to generate 
entanglement between two remote nodes in two different laboratories separated by a 
distance of 21 meters and linked by an optical fiber. 
They were able to maintain entanglement for about 100 microseconds while entanglement 
generation itself took about a single microsecond.

Later, a team from Technical University of Vienna~\cite{Vienna} succeeded in coupling Cesium 
atoms to an optical fiber and storing quantum information over a period of time that is
long enough to sustain entanglement over distances (hundreds of kilometers) that are large enough
to achieve reliable long distance communication.

The Vienna team extended coherence time to several milliseconds  
and given that speed of light in an optical fiber is about 200 kilometers per 
millisecond, a substantial separation increase is henceforth achievable potentially
reaching several hundred kilometers between nodes over which entanglement and 
coherence are maintained, paving the way to long-distance QC.

\subsubsection{Quantum noise}


At low temperature, very high frequency $hf >  k_B T$, mesoscopic scale or when considering
single carrier, quantum dot devices... quantum noise becomes larger than thermal 
implying a full reconsideration of traditional electronics that has long been
described by White (thermal noise with no relaxation time), Shot noise based on 
a single relaxation time (such as generation-recombination noise in semiconductors),
Pink noise ($1/f$) originating from a distribution of relaxation times...
 
Recently, entanglement has been shown to appear spontaneously in photon-assisted 
electrical noise occurring in quantum conductors consisting of an ac-biased tunnel 
junction  cooled at low temperature~\cite{Sherbrooke}. 

The experiments were performed in Sherbrooke~\cite{Sherbrooke} at 18 mK~\cite{field} on 
a Al/Al$_2$O$_3$/Al tunnel junction with resistance of 70 $\Omega$, 
the signal being emitted by the junction analyzed at two frequencies $f_1$=7 GHz and $f_2$=7.5 GHz.

The total voltage applied on the junction is given by $V_{dc}+V_{ac} \cos 2 \pi f_0 t$
with frequency $f_0=f_1+f_2$ chosen to produce optimal junction response as explained below.

Firstly, junction noise becomes photon-assisted because ac-biasing injects photons in the junction.

\begin{figure}[htbp]
  \centering
    \resizebox{60mm}{!}{\includegraphics[angle=0,clip=]{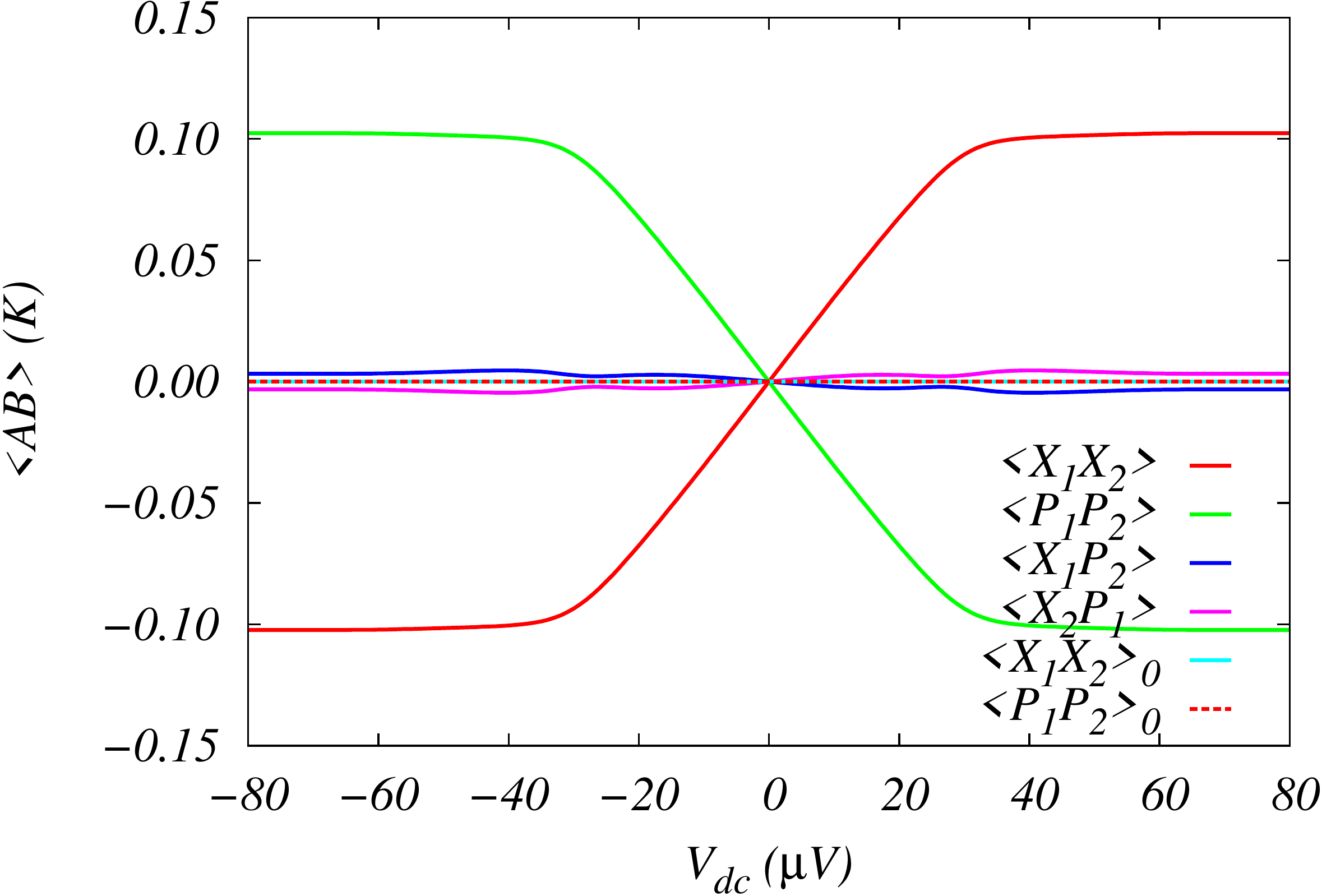}}  
\vspace*{-3mm} 
\caption{(Color on-line) Correlations in Kelvin units between currents at frequencies 
$f_1$ and $f_2$ versus $V_{dc}$ voltage applied to  the Al/Al$_2$O$_3$/Al tunnel junction for a fixed 
ac voltage $V_{ac}$= 37 $\mu$V. Violation of Bell inequalities is displayed only 
by $\langle X_1X_2\rangle$ and $\langle P_1P_2\rangle$ quantum correlations for non-zero 
applied ac voltage whereas all other correlators are null including $\langle X_1X_2\rangle_0$ 
and $\langle P_1P_2\rangle_0$ that are evaluated when the ac voltage is zero.}
\label{correlations}
\end{figure}

Secondly, statistical correlations between currents at $f_1$ and $f_2$ as a function of dc voltage
showed that photons generated in pairs in the junction are entangled since 
their correlations violate Bell inequalities~\cite{Bell} as discussed below.
Defining "position" $X_1,X_2$ and "momentum" operators $P_1, P_2$ from frequency dependent
current operators $I(\pm f_{1}), I(\pm f_{2})$ as:

\be
X_{1,2}=\frac{I(f_{1,2})+I(-f_{1,2})}{\sqrt{2}}, \, P_{1,2}=\frac{I(f_{1,2})-I(-f_{1,2})}{i\sqrt{2}}
\ee

we use the QFDT (see section~\ref{QFDT}) to evaluate the various quantum correlations versus
dc voltage $V_{dc}$ applied to  the Al/Al$_2$O$_3$/Al tunnel junction for a fixed ac 
voltage $V_{ac}$= 37 $\mu$V in fig.~\ref{correlations}. 
Violation of Bell inequalities displayed by $\langle X_1X_2\rangle$ and $\langle P_1P_2\rangle$  
for non-zero $V_{ac}$ indicate entanglement in contrast with the other correlators that
do not display any variation with $V_{dc}$. $\langle X_1X_2\rangle_0$ and $\langle P_1P_2\rangle_0$ that 
are evaluated for $V_{ac}=0$ do not show violation
of Bell inequalities indicating that it is $V_{ac}$ that induces quantum correlations thus
yielding a simple electrical control parameter for entanglement.

\section{Discussions and conclusions}
In this work, the main unifying thread for the description of fluctuations, noise
and noise-based communication is the ubiquitous presence of harmonic oscillators 
represented mostly by photons.

While secure classical noise-based communication uses spread-spectrum sequences,
secure QC based on QKD implemented with entanglement
ties communicating parties in a way such that any attempt by some
eavesdropper to intercept or interfere in the communication process is
immediately sensed and treated appropriately.

Entanglement may be done between quantum objects such as atoms, electrons, photons etc...
however the preferred information carrier is the photon and the entanglement 
that can be based on polarization, momentum, spatial mode or time can be sustained over 
very large distances as demonstrated by the Vienna experiment.

Heralded entanglement not necessitating a common ancestor has even been applied by the
same Garching~\cite{Ritter2} (MPQ) group to transfer a polarization qubit from a photon to 
a single atom with 39\% efficiency and perform the reverse process, that is from the atom 
to a given photon with an efficiency of 69\%, proving once again that a long-distance
QC network based on entangled photons is a serious contender for secure communication.

On the other hand, the Sherbrooke experiment shows that a major component 
of noise-based QC is built within quantum noise since entanglement is produced 
in quantum conductors by a simple electrical (ac voltage) control.

Even if presently such entanglement occurs at very low temperature (18 mK), 
the result is still important since that particular type of entanglement could be 
exploited after appropriate conditioning with quantum cryptography techniques in order
to  secure information transfer and communication. 


Presently several secure QC schemes not based on entanglement exist, moreover
some other protocols not relying on key generation and distribution have 
also been developed.

For instance QSDC (Quantum Secure Direct Communication) is a branch of QC 
in which the message is sent directly between remote users without 
generating a key to encrypt it.

Practicality and robustness of schemes used in QC for securing transmission
of information will finally decide which of the different methods and 
protocols will be ultimately adopted as reliable for secure mass communication.    


\appendix

\section{Stochastic processes and Noise}
A stochastic process is a random function of time that behaves in a way such 
that
every value $\xi$  it takes is distributed with a PDF $p(\xi;t)$:
\be
\langle \xi(t)\rangle= \int\limits_{-\infty}^{\infty} \xi p(\xi;t) d\xi
\ee
The ensemble~\cite{Gardiner} average $\langle ...\rangle$ is taken over the possible occurrences of 
the random function $\xi(t)$.

When the PDF is given by a Boltzmann distribution as illustrated in Section II,
the ensemble average is called thermal:

\be
\langle \xi(t)\rangle_{\beta}= A \int\limits_{-\infty}^{\infty} \xi e^{-\beta E} d\xi
\ee

with $A$ a normalization constant, $E$ the system energy and $\beta$ the inverse temperature. 

Time average of $\xi(t)$ is defined by:
\be
\overline{ \xi(t)}= \lim_{T \to +\infty}\frac{1}{T} \int\limits_{0}^{T} \xi(t) dt
\ee

In this work we assume the validity of the Ergodic theorem~\cite{Gardiner,Thermo} that ensures the
equivalence of time averaging and ensemble averaging.

The ensemble average  of the random function $\xi^2(t)$ is given by:
\be
\langle \xi^2(t)\rangle= \int\limits_{-\infty}^{\infty} \xi^2 p(\xi;t) d\xi
\ee 
whereas the mean-squared ensemble average is given by 
$\langle \delta \xi^2\rangle=\langle \xi^2(t)\rangle -{\langle \xi(t)\rangle}^2$.
The auto-correlation depending on two instants $t_1, t_2$ is:
\be
{\cal R}(t_1, t_2)= \langle \xi(t_1) \xi(t_2)\rangle = 
\int\limits_{-\infty}^{\infty}\int\limits_{-\infty}^{\infty}
\xi_1 \xi_2 p(\xi_1 ,\xi_2; t_1, t_2) d\xi_1 d\xi_2
\ee

with $p(\xi_1 ,\xi_2; t_1, t_2)$ representing the two times  
PDF for the distribution of the variables $\xi_1 ,\xi_2$.

Note that in the quantum case, ensemble average is replaced by quantum average.

For a stationary process we have the properties: $\langle \xi(t)\rangle$= constant,
and ${\cal R}(t_1, t_2)=R(t_1-t_2)$, with $R$ a function of a single argument
in contrast with ${\cal R}$.

Taking $t_1=t$ and $t_2=t+\tau$ we infer that ${\cal R}(t,t+\tau)= R(\tau)$ is a 
function of $\tau$ only meaning it is independent
of the instant $t$ at which we start observing the process and depends only of 
the interval of time $\tau$ during which it is being observed.

Actually we have a symmetry under the exchange of the time instants $t_1, t_2$
resulting into the property:

\be
{\cal R}(t_1, t_2)={\cal R}(t_2, t_1)=R(|t_1-t_2|) 
\ee

Moreover, we expect physically decorrelation of the variables $t_1,t_2$ as
the delay separating them increases, that is: $R(|t_1-t_2|)$ is a decreasing function
of its argument $|t_1-t_2|$.

The Ergodic theorem~\cite{Reif} specifies that ensemble average is equivalent
to time average in many systems. In glassy systems, ergodicity is not valid.

Stationary noise like any other stochastic process has an auto-correlation function
${\cal R}(t,t+\tau)$ function of  only $\tau$, 
hence we can use the definition $R(\tau)= {\cal R}(t,t+\tau)= {\cal R}(0,\tau)$.

This leads to classifying noise according to its Power Spectral Density (PSD) 
defined as the Fourier transform of the auto-correlation 
function $R(\tau)=\langle V(t) V(t+\tau)\rangle$ of the voltage fluctuations:
\be
S_V(f)=\int\limits_{-\infty}^{+\infty} \exp (-i2\pi f \tau) R(\tau) d\tau
\ee

Note that the integral prefactor is 1 whereas in the angular frequency
case $S_V(\omega)$, the prefactor is $2\pi$. \\
The inverse Fourier transform yields:
\be
R(\tau) =\int\limits_{-\infty}^{+\infty} \exp (i2\pi f \tau) S_V(f) df
\ee

The validity of the direct and inverse Fourier transforms is conditioned by 
the existence of the respective integrals:

\be
\int\limits_{-\infty}^{+\infty} |R(\tau)| d\tau < \infty, \hspace{1cm} 
\int\limits_{-\infty}^{+\infty}| S_V(f)| df < \infty
\ee

When $\tau=0$ we get $\langle V(t) V(t)\rangle =R(0) =\int\limits_{-\infty}^{+\infty} S_V(f) df$.\\

In the case of classical systems, the auto-correlation is always real and thus the
PSD is always symmetric with respect to the frequency. Consequently it is possible to rewrite 
the above formulae as:

\bea
S_V(f)=4 \int\limits_{0}^{+\infty} \cos (2\pi f \tau) R(\tau) d\tau, \nonumber \\
R(\tau) =\int\limits_{0}^{+\infty} \cos (2\pi f \tau) S_V(f) df  
\eea

The auto-correlation may also be written as $R(\tau) =\langle V(\tau) V(0)\rangle - \mean{V}^2$.
A simple classification of Noise is possible when the PSD behaves approximately 
as $S(f) \sim 1/f^{n}$. Consequently we have the following cases:
\begin {itemize}
\item Blue noise: $n=-1$ or  $S_V(f) \sim f $. This originates from the blue
glow observed in Cherenkov radiation emitted by a charged particle traveling 
in a dielectric at a velocity larger than light phase velocity in that medium.
\item White noise: $n=0$ or  $S_V(f) \sim $ constant. This is noise heard when a 
radio or TV broadcast station has stopped emitting.
\item Pink noise: $n=1$ or  $S_V(f) \sim 1/f$. $1/f$ noise is encountered in many 
areas of Science and Technology.
It is called also flicker noise and occurs also in music, earthquakes, 
floods... 
\item Brown noise: $n=2$ or  $S_V(f) \sim 1/f^2$. This is analogous to random 
walk (Brownian motion) and the
PSD falls faster than $1/f$.
\item Black noise: $n>2$ or  $S_V(f) \sim  1/f^{n}$. This is the opposite of 
white noise, in the sense that
the PSD falls off very quickly.

\end{itemize}

\section{The classical fluctuation-dissipation theorem (FDT)}

Einstein developed in 1905 a theory that explained Brownian motion on the 
basis of the kinetic theory of gases.

The 1D motion of the particle is essentially a random walk, 
with steps to right and left as equally probable. 

Einstein suggested that the mean kinetic energy per degree
of freedom of the particle should be given by statistical mechanics and the 
equipartition of energy as:
\be
\frac{1}{2} m \mean{v^2}=\frac{1}{2}k_B T
\ee
                                     
where $m$ is the mass of the particle, $v$ its instantaneous velocity 
component in the $x$-direction, the mean-squared value $\mean{v^2}$
is equal to the time average  $\mean{v^2}$ and Einstein did not make
a distinction between the two averages assuming Ergodicity.

The displacement $x$ in the $ {\bm x}$-direction
during the time interval $t$. Einstein showed that
\be
\mean{(x-\mean{x})^2} = 2 D t
\ee

with $D$ the diffusion constant of the particle.

Let us prove these results directly from a classical equation of motion
extended to comprise random excitation terms (Langevin equation):

\begin{equation}
m \frac{dv(t)}{dt}= -\alpha v + \xi(t)
\label{langevin3}
\end{equation}

The viscous term $\alpha v$ with coefficient $\alpha= 6 \pi \eta a$
is a Stokes term depending on an average damping coefficient $\eta$ on the particle from
irregular impacts on the particle with radius $a$ from the surrounding fluid.

$\xi(t)$ is a random term originating from
the surrounding medium considered as a reservoir at temperature $T$
with the following statistical properties: 

\begin{equation}
\mean{\xi(t)}=0 \mbox{  and  } \mean{\xi(t)\xi(t')} = \lambda \delta(t-t') 
\label{white}
\end{equation}

The above Langevin equation~\cite{Gardiner} contains
a time-dependent random excitation term $\xi(t)$ in an otherwise ordinary
differential equation (ODE) (see Section II).

Writing $\alpha= m \gamma$ we perform a direct integration of the first-order ODE:

\begin{equation}
v(t)= v_0 e^{-\gamma t} + \frac{1}{m} \int\limits_0^t \xi(t') e^{-\gamma (t-t')} dt'
\label{langevin4}
\end{equation}

with $v_0$ the initial velocity at $t=0$. The average of eq.~\ref{langevin2} yields 
$\mean{v(t)}= v_0 e^{-\gamma t}$ since $\mean{\xi(t)}$=0. This indicates that $\gamma$
is an inverse relaxation time of the initial velocity much like $1/RC$ in the circuit
encountered in Section II.

The auto-correlation of the velocity is given by:

\begin{equation}
\sigma_v^2(t) = \mean{v(t)v(t')}-\mean{v(t)}^2= \frac{\lambda}{m^2} \exp \left(-\gamma {|t-t'|}\right)
\label{langevin5}
\end{equation}

In order to get the value of $\lambda$ we recall Einstein (asymptotic) result
of Brownian motion.  

Setting $t=t'$ and identifying $\frac{\lambda}{m^2}$ with $\frac{D}{\gamma}$ we get
the value of $\lambda$.   Thus:

\begin{equation}
\sigma_v^2(t)= 2D \int\limits_0^t \xi(t') e^{-2\gamma (t-t')} dt'= \frac{D}{\gamma}(1-e^{-2\gamma t}),
\hspace{2mm} t > 0
\label{langevin6}
\end{equation}

This means velocity dispersion increases initially 
with time $\sigma_v^2(t)= 2D t$ when time $t \ll \tau_r$, to finally saturate at the
value $\sigma_v^2(t)\sim \frac{D}{\gamma}$ for time $t \gg \tau_r$.

Taking account of the average kinetic energy $\mean{E}=m \mean{v(t)^2}/2$ and
recalling the equipartition theorem $\mean{E}=k_B T/2$, we get $\gamma=\frac{m}{k_B T} D$.

Hence we can rewrite the auto-correlation formula as:

\begin{equation}
\gamma= \frac{1}{2m k_B T} \int \limits_{-\infty}^{\infty} \mean{\xi(t) \xi(t+\tau)} d\tau 
\hspace{2mm} t > 0
\label{langevin7}
\end{equation}

This is the classical fluctuation-dissipation theorem with auto-correlation representing
fluctuation and friction coefficient $\gamma$ representing dissipation.

\section{Density of states for particles and elementary excitations}
According to Kittel~\cite{Kittel} the density of states of solid-state excitations
$g(\omega)$ in $d$ dimensions for a system of typical linear length $\ell$ is given by 
$g(\omega)~=~\left(\frac{\ell}{2\pi}\right)^d~\int\frac{dS_{\omega}}{v_g}$ with ${\bf k}$ 
integration performed such that $\omega <\omega({\bf k})< \omega+d\omega$.
$\epsilon({\bf k})=\hbar \omega({\bf k})$ is the energy dispersion, $v_g$ 
is the group velocity modulus of the elementary excitations~\cite{Kittel}: 
$v_g=\frac{1}{\hbar}|\nabla_{\bf k} \epsilon({\bf k})|$ and $dS_{\omega}$ is the differential 
area element on the constant energy surface 
$\omega({\bf k})= \omega$. It is possible to generalize this formula to 
$g(\omega)~=~N_p~\left( \frac{\ell}{2\pi}\right)^d~\int~\frac{dS_{\omega}}{v_g}$
where $N_p$ is the number of excitation polarizations.

When the excitations are real particles (photons, electrons...) and possess a spin $S$,
 $N_p= 2 S+1$ when the
particles have non-zero mass (electrons) and $N_p=2$ for zero-mass particles (such as photons).
E. Wigner~\cite{Wigner} showed in 1939, on the basis of Lorenz invariance,
that the photon (or any other massless particle with spin $S$) moves with the velocity of light $c$, 
even in the center-of-mass frame. Exploiting rotational symmetry around $c$ direction yields
only two polarizations: left or right circular corresponding to $m_S=\pm S$ spin eigenstates.

In the case of elementary excitations  (phonons, plasmons, magnons, excitons...) $N_p=1$
regardless of the statistics. 

Specializing to the "Debye" case $\omega({\bf k})= v_g |{\bf k}|=v_g k$, the above expression
of the density of the states can be expressed analytically since the constant energy surface 
$\omega({\bf k})= \omega$ is a hypersphere with radius $k=\omega/v_g$ and surface
equal to $S_{\omega}={s}_1 k^{d-1}$ where ${s}_1$ is the unity radius hypersphere surface 
given by~\cite{Reif} ${s}_1= 2 \pi^{d/2}/\Gamma(d/2)$. \\
Collecting expressions we get $g(\omega)~=~N_p\left(\frac{\ell}{2\pi}\right)^d~\frac{S_{\omega}}{v_g}$ 
thus $g(\omega)~=~N_p\left(\frac{\ell}{2\pi}\right)^d~\frac{1}{v_g}\frac{2 \pi^{d/2}}{\Gamma(d/2)}\left(\frac{\omega}{v_g}\right)^{d-1} $

In an oscillator, excitations are quantized with average energy for $n$ quanta as 
$\hbar \omega (\mean{n}+\frac{1}{2})$ in the interval 
$[\omega, \omega  + d\omega]$.
Multiplying the mean energy by the number of quanta (modes)
$g(\omega)d\omega$ in this interval yields:
$g(\omega) \hbar \omega (\mean{n}+\frac{1}{2}) d\omega$.

\section{Entropy hierarchy}
In order to establish a hierarchy of information entropies, recall that Shannon 
entropy is defined (in the discrete probability case) by:
\be
H_S=-\sum_{i=1}^N p_i \log_2 p_i
\ee
with $p_i$ the probability of occurrence of symbol $i$ and $N$ the total number of symbols.

Guided by Hamming distance well-known in coding theory~\cite{Carlson}, one might draw an analogy between information and distance in order to establish a hierarchy of entropies.

For any vector $\bm{x}$ with components $x_i, i=1,..N$, its distance from origin 
or norm is defined according to the following $\ell_p$-norm formula:
\be
||\bm{x}||_p=\left(\sum_{i=1}^N |x_i|^p \right)^{\frac{1}{p}}
\ee

Thus: $||\bm{x}||_1=\sum_{i=1}^N |x_i|$ is the $\ell_1$-norm, whereas the ordinary Euclidean
norm is $||\bm{x}||_2=\left(\sum_{i=1}^N |x_i|^2 \right)^{\frac{1}{2}}$
the $\ell_2$-norm. Taking the case $p \rightarrow \infty$ we get the
Infinity norm $\ell_\infty$ with $||\bm{x}||_\infty=\max_{i=1...N} |x_i|$.
Mathematically, the three norms are equivalent, however computationally wise, 
$\ell_\infty$ norm is the most efficient in terms of number of arithmetic operations.

In the entropy case, we define the order $q \in [0,\infty[$ R\'enyi function:

\be
R_q=\frac{1}{1-q} \log_2 \left(\sum_{i=1}^N p_i^q \right)
\ee

The R\'enyi entropy is additive like Shannon's and for $q \rightarrow 1$, they are same:

\bea
R_1=\lim_{q \rightarrow 1} R_q &=& -\frac{d}{dq}{\left(\log_2 \left(\sum_{i=1}^N p_i^q\right)\right)}_{q=1}  \nonumber \\
   &=&-\sum_{i=1}^N p_i \log_2 p_i \equiv H_S
\eea

Using the analogy with the Infinity norm $\ell_\infty$, we obtain the min-entropy as:

\be
R_\infty=-\max_{i=1...N} \log_2 p_i
\ee

Similarly to the $\ell_\infty$ norm, the R\'enyi entropy is very efficient, 
computationally wise.

Considering for an example, Poisson distributed photons~\cite{Poisson},   
with mean $\mean{n}$ and probabilities $p_i= \frac{e^{-\mean{n}} \mean{n}^i}{i!}$, 
Shannon entropy is given by:
\be
H_S=\frac{1}{2} \log_2(2 \pi e \mean{n})
\ee

In comparison, the min-entropy is obtained after estimating $\max_{i=1...N} \{p_i\}$ 
and that occurs when $i \approx \lfloor\mean{n}\rfloor$, the integer part of $\mean{n}$, thus:

\be
R_\infty=-\log_2 [ \frac{ e^{-\mean{n}} \mean{n}^{\lfloor\mean{n}\rfloor} }{\lfloor\mean{n}\rfloor!}  ]
\label{Renyi}
\ee
 
Entropy hierarchy is represented by the inequalities 
$0 \le R_\infty \le R_q \le H_S \le R_{q'}$ (with $q>1$ and $0<q'<1$). 
In the case of perfectly random $n-$bit strings, both entropies $H_S$ and $ R_\infty$ per bit 
are equal to one, since $p_i=1/N, \forall i$ with $N=2^n$ and $\max_{i=1...N} \{p_i\}=1/N$.


\begin{thebibliography}{99}
\bibitem{Duality} Wave-particle duality applied to massless photons is due,
surprisingly, to Einstein. De Broglie used duality for massive particles and 
considered that the photon had a finite mass.
\bibitem{factor2} The factor 2 discrepancy ($4R k_BT$ versus $2R k_BT$) occurring 
in the PSD $S_V(\omega)$ or in the voltage fluctuation $\sigma_V^2$ is resolved by 
multiplying by 2 the contributions for positive and negative $\omega$.
\bibitem{classical} In modern statistical physics language, the classical limit is obtained
when the Bose-Einstein factor $\mean{n}=\frac{1}{e^{\beta (\hbar \omega- \mu)} -1} \ll 1$
with $\mu$ the chemical potential or when mean particle distance is much larger than
thermal wavelength~\cite{Landau}. 
When $\beta \mu \rightarrow -\infty$, we recover
Boltzmann distribution $\mean{n} \approx  e^{-\beta (\hbar \omega- \mu)}$.
The photon number not being conserved, $\mu=0$ for all temperatures, implying that the photon
character is always quantum and the "classical" limit $k_B T \gg \hbar\omega$ invoked by Planck
bypasses the quantum nature of the photon.
\bibitem{electromag} The classical contribution to energy from the equipartition theorem
is $ 2 \times 2 \times \frac{1}{2} k_B T= 2 k_B T$ since the electromagnetic 
oscillator energy per mode is quadratic with respect to two 
orthogonal field components (electric and magnetic) 
and two polarizations (right and left circular) are present. 
The polarization contribution is in the prefactor 
$ 2 \times \frac{\omega^2}{\pi^2 c^3}$ which is same for both
Rayleigh-Jeans and Planck expressions of the photon spectral density.
\bibitem{continuous} Continuous variable quantum information processing corresponds to the quantum
version of the analog computer. A simple representative variable is $\ket{x}$ 
the position representation eigenstate, a continuous version of the qubit that belongs
to an infinite dimensional Hilbert space.
\bibitem{spinor} Mathematically, a spinor reverses its sign when the polar angle 
is rotated by 2$\pi$ since $\ket{\psi(\theta+2\pi,\phi)}=\hat{R}_z(\theta+2\pi,\phi)\ket{\psi(\theta,\phi)}
=-\ket{\psi(\theta,\phi)}$
\bibitem{field} In the low-temperature experiment performed at 18 mK, 
the junction is subjected
to a magnetic field forbidding Aluminum to become superconducting: Al has 
0.01 T critical field and 1.2 K critical temperature.
\end{thebibliography}
\end{document}